\documentclass{ieeeaccess}
\usepackage{hyperref}
\usepackage{lmodern}
\usepackage{pifont}
\usepackage{amsmath,amssymb,amsfonts}
\usepackage{graphicx}
\usepackage{multirow,multicol}
\usepackage{algorithm,algorithmic}
\usepackage{subcaption}
\usepackage{booktabs}
\usepackage{color,soul}

\soulregister\cite7
\soulregister\ref7
\soulregister\pageref7

\def\BibTeX{{\rm B\kern-.05em{\sc i\kern-.025em b}\kern-.08em
 T\kern-.1667em\lower.7ex\hbox{E}\kern-.125emX}}
\begin{document}
\history{Date of publication xxxx 00, 0000, date of current version xxxx 00, 0000.}
\doi{10.1109/ACCESS.2017.DOI}

%this title is better than the previous one!
\title{Privacy Preservation and Identity Tracing Prevention in AI-Driven Eye Tracking for Interactive Learning Environments}

\author{
\uppercase{Abdul Rehman}\authorrefmark{1},
\uppercase{Are Dæhlen}\authorrefmark{1},
\uppercase{Ilona Heldal}\authorrefmark{1},
\uppercase{Jerry Chun-Wei Lin}\authorrefmark{1},
}

\address[1]{Department of Computer Science, Electrical Engineering and
Mathematical Sciences, Western Norway University of Applied Sciences,
Bergen, Norway}

\markboth
{Author \headeretal: Preparation of Papers for IEEE TRANSACTIONS and JOURNALS}
{Author \headeretal: Preparation of Papers for IEEE TRANSACTIONS and JOURNALS}

\corresp{Corresponding author: arj@hvl.no, jerry.chun-wei.lin@hvl.no}

\tfootnote{}
\begin{abstract}
Eye-tracking technology provides an estimate of where a person is looking and is essential for understanding individual concentration and the psychological functions within the cognitive system. It can aid in understanding neurodevelopmental disorders and tracing a person's identity. However, this technology poses a significant risk to privacy, as it captures sensitive information about individuals and increases the likelihood that data can be traced back to them. 
This paper proposes a human-centered framework designed to prevent identity backtracking while preserving the pedagogical benefits of AI-powered eye tracking in interactive learning environments. We explore how real-time data anonymization, ethical design principles, and regulatory compliance (such as GDPR) can be integrated to build trust and transparency. Our approach aims to strike a balance between technological innovation and privacy preservation, providing insights into the design of responsible AI in educational contexts.
We first demonstrate the potential for backtracking student IDs and diagnoses in various scenarios using serious game-based eye-tracking data. We then provide a two-stage privacy-preserving framework that prevents participants from being tracked while still enabling diagnostic classification. 
The first phase covers four scenarios: 
I) Predicting disorder diagnoses based on different game levels. 
II) Predicting student IDs based on different game levels. 
III) Predicting student IDs based on randomized data. 
IV) Utilizing K-Means for out-of-sample data. 
In the second phase, we present a two-stage framework that preserves privacy. We also employ Federated Learning (FL) across multiple clients, incorporating a secure identity management system with dummy IDs and administrator-only access controls.
In the first phase, the proposed framework achieved 99.3\% accuracy for scenario 1, 63\% accuracy for scenario 2, and 99.7\% accuracy for scenario 3, successfully identifying and assigning a new student ID in scenario 4. In phase 2, we effectively prevented backtracking and established a secure identity management system with dummy IDs and administrator-only access controls, achieving an overall accuracy of 99.40\%. The proposed framework successfully secures students' data and prevents identity backtracking.
\end{abstract}
\titlepgskip=-15pt
\maketitle 
\IEEEpeerreviewmaketitle
\section{Introduction}
\label{intro}
\begin{figure}[!ht]
 \centering
 \includegraphics[width=\linewidth]{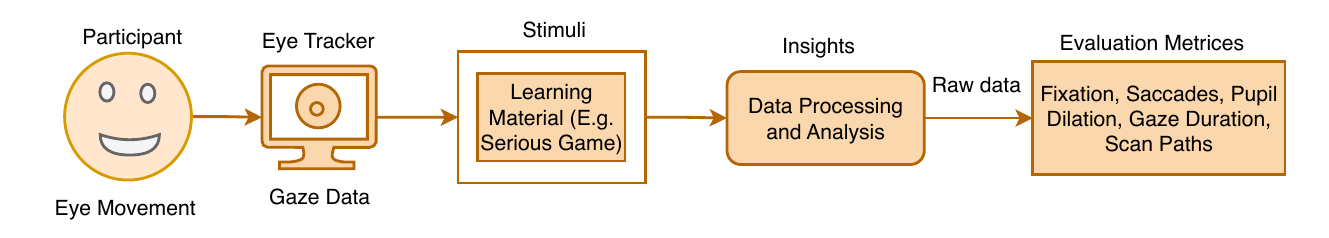}
 \caption{Process of Eye Tracking Data Collection}
 \label{fig1P}
\end{figure}

Eye-tracking technology provides an estimation of where a person is looking and is essential for understanding individual attention and the psychological functions of the cognitive system in the interactive learning system. Due to this distinctive attribute, eye tracking is a vital building block, allowing for various ubiquitous and interactive applications~\cite{khamis2018past}. The quick and precise tracking of an eye that efficiently covers the sensor field of view (also referred to as "near-eye" tracking) is advantageous for these and other applications, including laser eye surgery~\cite{neuhann2010static}. Fig. \ref{fig1P} shows how eye-tracking data is collected and processed from students. The students view learning materials while an eye tracker captures their gaze data, which is then processed to generate research insights. The figure includes key metrics, such as fixation, gaze duration, scan paths, and pupil dilation, which are typically measured in educational eye tracking studies~\cite{daehlen2024towards}. Analysis and interpretation of eye-tracking data is essential for studies that examine the cognitive processes of their participants~\cite{holmqvist2011eye}. Some analysis approaches are better suited for specific research objectives and generally vary from study to study, depending on task dependency and the rigidity of the stimuli. Analysis methods such as scan paths and heat maps perform longitudinal analysis for an entire data collection session~\cite{ScanPathsUsage}, while the area of interest analysis provides stimulus-specific insight~\cite{AOIUsage}. Examining raw gaze can be utilised for static task structures~\cite{CandLook}, along with processing additional eye movement features such as microsaccades and post-saccadic oscillations~\cite{alexander2020microsaccades}.

As artificial intelligence (AI) systems become increasingly integrated into educational technologies, eye-tracking is emerging as a powerful tool for enhancing interactivity and understanding learner behavior \cite{guilherme2019ai}. These systems offer opportunities for real-time feedback, personalized learning pathways, and improved engagement in digital learning environments. However, the deployment of such technologies also raises significant ethical and societal concerns, particularly around surveillance, data privacy, and identity inference \cite{farina2024ai}. The collection and analysis of visual attention data can inadvertently reveal sensitive personal attributes, including cognitive states, emotional responses, diagnosis (if labeled) and even user identity. Without appropriate safeguards, these systems risk becoming tools of unintentional surveillance, undermining trust and infringing on learners’ rights to privacy and autonomy. The tension between leveraging AI for educational enhancement and preserving individual privacy calls for a careful, human-centered design approach. 

\begin{figure}[!ht]
 \centering
\includegraphics[width=0.8\linewidth]{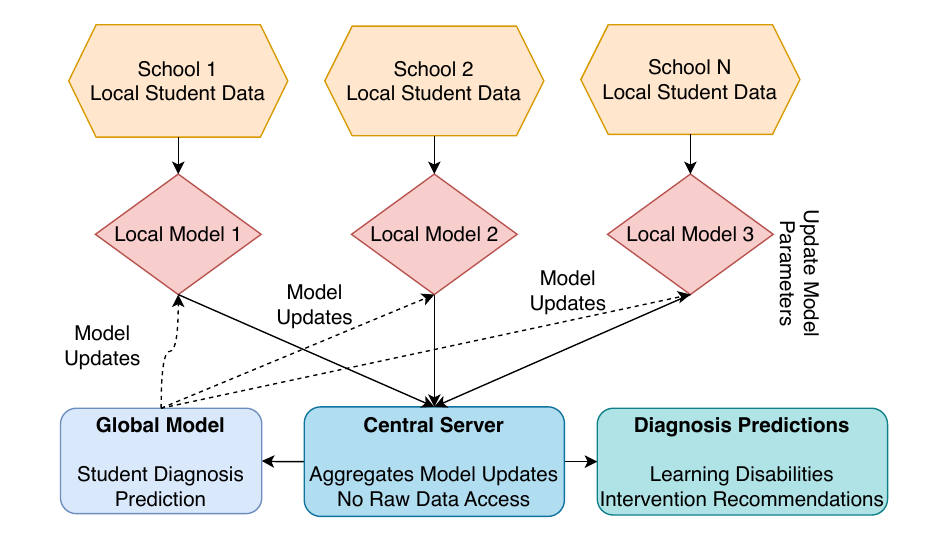}
 \caption{Student Diagnosis Prediction using FL}
 \label{fig2}
\end{figure}

Numerous studies have demonstrated that improvements in data linkage algorithms can re-identify individuals~\cite{david2022providing,na2018feasibility}. Backtracking from gaze data can lead to backtracking identity, disease or even sexual orientation~\cite{david2022providing}. Therefore, preventing this backtracking and preserving students' privacy is essential. The potential for a standardised ML model to fail to preserve privacy and data security is highlighted~\cite{papernot2018sok}. A suitable approach to overcoming these challenges is FL, which permits decentralised ML while protecting user privacy~\cite{lian2022layer,lian2022deep}. FL ensures secure and individualized healthcare solutions for people with neurodevelopmental disorders by facilitating collaborative learning across multiple devices without data centralization~\cite{mcmahan2017communication}. Eye-tracking and FL can be combined to improve remote monitoring and intervention techniques by enabling early and precise detection while protecting privacy. Fig. \ref{fig2} illustrates a scenario involving multiple schools, demonstrating how they can collaborate to develop a predictive model without sharing sensitive student data. Each school trains a local model with its local student data, and only model updates (not raw data) are sent to a central server. The server aggregates these updates to improve a global model that predicts the student's diagnosis and then distributes it back to the local models of the schools. This approach maintains student privacy while enabling collaborative diagnosis, prediction, and intervention recommendations across different schools.

This paper makes the following contributions:

\begin{itemize}
\item We first proposed a framework to backtrack the IDs and diagnoses in 4 scenarios: I) How can we predict the disorder diagnosis based on different game levels? II) How can we predict student IDs based on different game levels? III) How can we predict student IDs based on randomised data? IV) How can we use K-Means for Out of Sample Data (treating the first 8 students as existing and the 9th as new)? Next, it provides a two-stage privacy-preserving framework that prevents participants from being tracked while enabling diagnostic classification. Next, we utilize FL across multiple clients, employing a secure identity management system with dummy IDs and administrator-only access controls.

\item For the first phase, the proposed framework achieved 99.3\% accuracy using a decision tree for scenario 1; for scenario 2, it achieved an accuracy of 63\% using Random Forest; for scenario 3, it achieved an accuracy of 99.7\% using a decision tree, for scenario 4, it successfully identified and assigned a new student ID.
\item For the phase 2, it successfully prevented backtracking and secured an identity management system with dummy IDs and administrator-only access controls and achieved an overall accuracy of 99.40\%. 
\item The results reveal that the proposed framework works well for backtracking and subsequently helps secure students' data and prevent backtracking.
\end{itemize}

The paper is organized as follows: Section \ref{RW} presents the related work on machine learning and federated learning (FL) for NDD diagnosis and privacy preservation. Section \ref{PF} provides the proposed NDD diagnosis and privacy preservation framework. Section \ref{RA} presents the results and analysis. Section \ref{ethical} discusses the ethical considerations. Finally, Section \ref{conclusion} concludes the paper.

\section{Related Work}
\label{RW}
Kim et al.~\cite{kim2024development} aimed to create a DL model based on eye tracking and the RCFT to identify individuals with mental diseases who had impaired the encoding of visuospatial memory. Eye movements during a 3-minute RCFT memorization test were recorded to assess the structure and retention of psychosis, OCD, and healthy controls. These scores and the fixation points, which indicate areas of eye focus, were used to build a Long Short-Term Memory (LSTM) model with an attention mechanism that could differentiate between normal and impaired executive function. Du et al.~\cite{du2024privategaze} introduced PrivateGaze, which safeguards users' private data while interacting with black-box gaze tracking data, while maintaining estimation accuracy. PrivateGaze can effectively preserve users' data, such as identity and gender, as demonstrated by experiments on four benchmark datasets.

Perkovich et al.~\cite{perkovich2024conducting} used eye-tracking data of 41 autistic kids and 17 kids who do not have visual or auditory abnormalities but are at a higher risk of being diagnosed with autism in the future. Data collection among children with autism was 92.68\% successful. Islam et al.~\cite{islam2025involution} applied a hybrid involution-convolution strategy since eye-tracking signals mostly rely on spatial information, and they applied two image-processing techniques to detect them. With a size of just 1.36 megabytes, the suggested model yielded high accuracy. Yoo et al.~\cite{yoo2024development} aimed to create a screening model for ADHD by utilising eye-tracking characteristics from activities that represent neuropsychological abnormalities in ADHD, along with machine learning (ML). An effective model for classifying ADHD was then constructed to identify pertinent eye-tracking variables for ADHD. Eye-tracking factors were used to identify ADHD with reasonable accuracy.

\subsection{Machine and Deep Learning for Student Privacy}
Price et al.~\cite{price2019privacy} argued that, although big data enables efficient analysis of large and diverse datasets, it still poses significant privacy risks even when de-identification techniques are applied. Numerous studies have demonstrated that improvements in data linkage algorithms can lead to the re-identification of individuals. For example, a study by Na et al.~\cite{na2018feasibility} used machine learning algorithms (i.e., Random Forest) to re-identify 85.6\% of adults and 69.8\% of children in a physical activity cohort study after identifiers of protected health information were removed. Revathi et al.~\cite{revathi2024smart} used an ML-assisted method to protect and monitor children's health information in a blockchain setting. First, public data sources are utilized in the Hybrid Encryption Technique (HET) to gather the children's medical records. Elliptical curve cryptography (ECC) and attribute-based encryption (ABE) are two methods implemented in the HET. Conditional Random Fields (CRF) and Multi-Layer Perceptrons (MLP) are used in the Adaptive Machine Learning (HAML) technique. The findings suggest that the proposed approach outperforms traditional methods. Xei et al.~\cite{xie2022private} address these two concerns and provide students with a time-efficient and privacy-conscious anomaly detection solution for using wearable sensors in a mobile cloud computing environment. The high volume and regular updates of health data collected through multiple wearable sensors and transmitted to the cloud platform cause low anomaly detection performance. Furthermore, integrating student health data into cloud platforms can be challenging due to its sensitive nature.

\subsection{Federated Learning for Healthcare Application}
Electronic health records (EHRs), wearable technology, and Internet of Things (IoT) devices are all combined in healthcare systems to provide a networked system for ongoing patient monitoring. Large-scale data, including neurological, biological measurements, psychological patterns, and ecological data, are being created due to the expansion of IoT in health maintenance; these data have great significance for machine-learning models in patient-centred care~\cite{antunes2022federated}. By enabling ML models to train on a variety of data sources without the need for a centralised database, FL supports this approach. Since data accessibility is crucial in public health contexts, this is especially crucial for clinics and companies that operate across legal systems with disparate data governance rules. 

These challenges necessitate the development of more effective security solutions for utilizing FL in healthcare applications, such as clinical decision support systems, medical imaging evaluation, patient risk forecasting, remote tracking, personalized healthcare, and electronic health record optimization. This can ensure that model upgrades are free of contaminants from deceptive input attacks, corrupted input exploitation, or training data recovery~\cite{manzoor2024survey}. Therefore, FL is crucial for protecting sensitive data in healthcare IoT networks and fostering a sense of security in AI applications for patients and healthcare practitioners~\cite{alsamhi2024federated}. The DEEP-FEL federated edge learning system, as proposed by Lian et al.~\cite{lian2022deep}, enabled medical devices in numerous institutions to collaborate on training a global model without sharing raw data. A hierarchical ring topology must first be established to reduce the centralization of the traditional training framework. The formation of the ring is then treated as an optimisation issue that can be resolved using a trustworthy heuristic approach. Hanif et al.~\cite{hanif2022federated} extracted inter-institutional diagnostic patterns from disease epidemiology in retinopathy of prematurity (ROP) by using FL, a method for collaboratively training deep learning (DL) algorithms without revealing patient records. In the newborn critical care units of seven different hospitals, 5,245 retinal images were obtained from patient examinations. DL models are trained for disease classification using either an FL approach or a centralized multi-institutional dataset with clinical labels.

Overall, some studies have focused on providing privacy and security for user data; however, considerable work remains to address the limitations of backtracking risk and eye-tracking application-oriented use cases that we aim to solve in this paper.

\section{Problem Definition / Mathematical Formulation}
\begin{itemize}
\item Let $\mathcal{D} = \{(x_i, y_i, l_i, s_i)\}_{i=1}^N$ denote the overall dataset comprising $N$ labeled records.
\item Each $x_i \in \mathcal{D}$ represents an attribute/feature capturing gaze-related characters or behavior for the $i$-th record.
\item The corresponding label $y_i \in \mathcal{Y}$ indicates the target output class to be predicted, such as a diagnostic disorder label or student ID.
\item The variable $l_i \in \{1, 2, 3\}$ specifies the cognitive difficulty level of the task associated with sample $i$.
\item $s_i \in \{1, 2, \ldots, S\}$ denotes the unique student ID for the $i$-th record.
\item $N$ refers to the total number of samples in the dataset.
\item The predictive model is defined as a function $f_\theta: \mathbb{R}^d \rightarrow \mathcal{Y}$ parameterized by $\theta$ where $\mathbb{R}^d$ represents vector with d real-valued entries.
\end{itemize}

Below is a mathematical formulation for both phases.
\subsection{Phase 1}
\subsubsection{Scenario 1}
In this setup, we investigate the model's ability to generalize across task complexity by training it on gaze data obtained from tasks of lower difficulty and evaluating its performance on data from higher-complexity tasks, shown in Eq.~(\ref{eq.1}) and Eq.~(\ref{eq.2}).

\begin{align}
\mathcal{D}_{\text{train}} &= \{(x_i, y_i, l_i, s_i) \in \mathcal{D} \mid l_i \in \{1, 2\}\} 
\label{eq.1} \\
\mathcal{D}_{\text{test}} &= \{(x_i, y_i, l_i, s_i) \in \mathcal{D} \mid l_i = 3\}
\label{eq.2}
\end{align}

The objective is to determine the optimal model parameters $\theta^*$ by minimizing the loss function $\mathcal{L}_{f}$ over the training set shown in Eq.~(\ref{eq.3}):

\begin{equation}
\theta^* = \arg\min_{\theta} \sum_{(x_i, y_i) \in \mathcal{D}_{\text{train}}} \mathcal{L}_{f}(f_\theta(x_i), y_i)
\label{eq.3}
\end{equation}

After training, the model is evaluated using data from level 3 tasks, and its prediction accuracy is computed as shown in Eq.~(\ref{eq.5}):

\begin{equation}
\text{Model Performance} = \frac{1}{|\mathcal{D}_{\text{test}}|} \sum_{(x_i, y_i) \in \mathcal{D}_{\text{test}}} \mathbb{I}(\hat{y}_i = y_i)
\label{eq.5}
\end{equation}

Here, $\hat{y}_i = \arg\max_{y \in \mathcal{Y}} f_\theta(x_i)$ denotes the predicted label, and $\mathbb{I}$ is the indicator function, returning 1 when the prediction matches the ground truth. This cross-task evaluation strategy examines whether gaze-based diagnostic patterns remain stable when cognitive task complexity increases, a crucial criterion for developing generalizable and reliable diagnostic frameworks.

\subsubsection{Scenario 2}
In this scenario, we follow a setup similar to Scenario 1 by dividing the dataset according to task difficulty levels. However, instead of predicting disorder labels, the objective here is to identify individual students based on their unique gaze signatures. The training set consists of data from tasks of lower difficulty ($l_i \in \{1, 2\}$), while testing is conducted exclusively on level 3 data shown in Eq.~(\ref{eq.6}), (\ref{eq.6a}):

\begin{align}
\label{eq.6}
\mathcal{D}_{\text{train}} &= \{(x_i, s_i, l_i) \in \mathcal{D} \mid l_i \in \{1, 2\}\} \\
\mathcal{D}_{\text{test}} &= \{(x_i, s_i, l_i) \in \mathcal{D} \mid l_i = 3\} \label{eq.6a}
\end{align}

Given the multiclass nature of this identification task, we optimise the model using the categorical cross-entropy loss function shown in Eq.~(\ref{eq.7}):

\begin{equation}
\mathcal{L}(f_\theta(x_i), s_i) = -\sum_{k=1}^{S} \mathbb{I}(s_i = k) \log(f_\theta(x_i)_k) 
\label{eq.7}
\end{equation}

Here, $f_\theta(x_i)_k$ represents the predicted probability of assigning input $x_i$ to student $k$. Model performance is quantified using classification accuracy on the testing set, shown in Eq.~(\ref{eq.8}):

\begin{equation}
\text{Accuracy} = \frac{1}{|\mathcal{D}_{\text{test}}|} \sum_{(x_i, s_i) \in \mathcal{D}_{\text{test}}} \mathbb{I}(\hat{s}_i = s_i) 
\label{eq.8}
\end{equation}

where the predicted student identity is given by $\hat{s}_i = \arg\max_{k \in \{1,2,\ldots,S\}} f_\theta(x_i)_k$. This evaluation set is designed to test the stability of student-specific gaze patterns under increased task complexity, offering insights into the potential of gaze data for biometric identification in cognitively dynamic environments.

\subsubsection{Scenario 3}
In this scenario, a baseline machine learning setup, we apply a conventional strategy by randomly partitioning the complete dataset into training and testing subsets. Specifically, 60\% of the data is used for training, and the remaining 40\% is held out for evaluation. The equation used for training and testing is shown in Eq.~(\ref{eq.9}), (\ref{eq.9a}). We employ stratified sampling to preserve the overall distribution of class labels within both subsets. This ensures that the proportion of samples from each class $c \in \mathcal{y} (labels)$ remains approximately consistent between the whole dataset and the training set.

\begin{align}\label{eq.9}
\mathcal{D}_{\text{train}} &\subset \mathcal{D}, \quad |\mathcal{D}_{\text{train}}| = 0.6 \cdot |\mathcal{D}|
 \\
\mathcal{D}_{\text{test}} &= \mathcal{D} \setminus \mathcal{D}_{\text{train}}, \quad |\mathcal{D}_{\text{test}}| = 0.4 \cdot |\mathcal{D}|
\label{eq.9a}
\end{align}

Furthermore, we incorporate $K$-fold cross-validation\footnote{\url{https://scikit-learn.org/stable/modules/cross_validation.html}} to enhance robustness in performance estimation. The dataset is split into $K$ non-overlapping and mutually exclusive subsets (folds), as shown in Eq.~(\ref{eq.11}) and Eq.~(\ref{eq.12}):

\begin{align}
\mathcal{D} &= \bigcup_{k=1}^{K} \mathcal{D}_k
\label{eq.11} \\
\mathcal{D}_i \cap \mathcal{D}_j &= \emptyset \quad \text{for } i \neq j 
\label{eq.12}
\end{align}

The average performance across all $K$ folds is computed using Eq.~(\ref{eq.13}):

\begin{equation}
\text{CV-Performance Evaluation} = \frac{1}{K} \sum_{k=1}^{K} \text{Performance}_k 
\label{eq.13}
\end{equation}

where $\text{Performance}_k$ denotes the evaluation metric when the model is trained on $\mathcal{D} \setminus \mathcal{D}_k$ and tested on $\mathcal{D}_k$. This random-split evaluation is a standard benchmark, enabling direct comparison with the more cognitively motivated, level-based splitting strategies discussed in previous scenarios.

\subsubsection{Scenario 4}
In this scenario, we explore an unsupervised strategy to detect the presence of new, previously unseen students by leveraging clustering-based methods. The dataset is split into two disjoint subsets: one containing data from existing students and another comprising samples from a potential new individual, as shown in Eqs.~(\ref{eq.14} and (\ref{eq.15}):

\begin{align}
\mathcal{D}_{\text{existing}} &= \{(x_i, s_i) \in \mathcal{D} \mid s_i \in \{1, 2, \ldots, 8\}\} 
\label{eq.14} \\
\mathcal{D}_{\text{new}} &= \{(x_i, s_i) \in \mathcal{D} \mid s_i = 9\} 
\label{eq.15}
\end{align}

To model the structure of existing students, we apply K-Means clustering on $\mathcal{D}_{\text{existing}}$ and derive $K$ representative clusters using the Elbow method \footnote{\url{https://www.geeksforgeeks.org/machine-learning/elbow-method-for-optimal-value-of-k-in-kmeans/}}. For each cluster $C_j$, we compute its centroid $\mu_j$ using the expression shown in in Eq.~(\ref{eq.16}):

\begin{equation}
\mu_j = \frac{1}{|C_j|} \sum_{x_i \in C_j} x_i
\label{eq.16}
\end{equation}

When a new gaze data ($N_{gz}$) sample $x_{N_{gz}} \in \mathcal{D}_{\text{new}}$ is encountered, it is provisionally assigned to the nearest cluster based on the minimum squared Euclidean distance \cite{liberti2014euclidean} shown in Eq.~(\ref{eq.17}):

\begin{equation}
j^* = \arg\min_{j \in \{1, 2, \ldots, K\}} \|x_{N_{gz}} - \mu_j\|_2^2 \label{eq.17}
\end{equation}

To assess whether the sample truly belongs to one of the known clusters or represents a novel case, we define a novelty score ($N_S$) shown in Eq.~(\ref{eq.18}):

\begin{equation}
N_S(x_{\text{new}}) = \min_{j \in \{1, 2, \ldots, K\}} \|x_{\text{new}} - \mu_j\|_2 
\label{eq.18}
\end{equation}

If this $N_S$ exceeds a predefined threshold $\tau$ shown in Eq.~(\ref{eq.19}), the sample is marked as potentially belonging to a previously unseen identity:

\begin{equation}
\text{Novelty}(x_{N_{gz}}) > \tau 
\label{eq.19}
\end{equation}

To select an appropriate number of clusters $K$, we employ the silhouette coefficient \footnote{\url{https://scikit-learn.org/stable/modules/generated/sklearn.metrics.silhouette_score.html}} shown in Eq.~(\ref{eq.20}) as a cluster quality metric:

\begin{equation}
S(i) = \frac{b(i) - a(i)}{\max\{a(i), b(i)\}} 
\label{eq.20}
\end{equation}

Here, $ an (i)$ denotes the average intra-cluster distance for point $i$, and $b(i)$ represents the lowest average distance between point $i$ and all samples in the nearest neighbouring cluster. A higher silhouette score indicates more well-separated and coherent clusters, facilitating more reliable outlier detection for new student identification.

\subsection{Phase 2}
\subsubsection{Dual Layer Protection}

To protect student identities, the framework implements a dual-layer identification mechanism.

\textbf{Layer 1: Temporary Dummy IDs: } 
Each true student identifier $s_i$ is obfuscated using a time-sensitive transformation function, generating a temporary dummy ID $d_i$ as defined in Eq.~(\ref{eq.21}):

\begin{equation}
d_i = M_t(s_i, Ek_1) 
\label{eq.21}
\end{equation}

Here, $Ek_1$ is a secure encryption key, and $t$ represents the current period. The mapping function $M_t(\cdot)$ is periodically updated to ensure the short-lived validity of dummy identifiers shown in Eq.~(\ref{eq.22}):

\begin{equation}
M_{t+\Delta} \neq M_t \quad \text{for some interval } \Delta
\label{eq.22}
\end{equation}

\textbf{Layer 2: Admin-Level Recovery: } 
Authorised administrators can reverse the mapping using an inverse transformation with an additional secure key $Ek_2$, as described in Eq.~(\ref{eq.23}):

\begin{equation}
s_i = M_t^{-1}(d_i, Ek_1, Ek_2)
\label{eq.23}
\end{equation}

This layered scheme ensures that operational components handle only anonymised IDs, while true identities remain accessible only through restricted administrative functions.

\subsubsection{Privacy-Preservation}
To further preserve user privacy, the model is trained using an FL approach, where each student's device retains its local dataset $\mathcal{D}_s = \{(x_i, y_i) \mid s_i = s\}$ without sharing raw data.

The global model $\theta^t$ is updated iteratively across devices through the following protocol:

\begin{enumerate}
 \item The central server dispatches the current global model parameters $\theta^t$ to each participating device.
 
 \item Each device $s$ performs local training and computes an updated model $\theta_s^{t+1}$ using stochastic gradient descent as described in Eq.~(\ref{eq.24}):

 \begin{equation}
 \theta_s^{t+1} = \theta^t - \eta \nabla \mathcal{L}_s(\theta^t) \label{eq.24}
 \end{equation}

 \item Devices transmit only the parameter updates to the server as described in Eq.~(\ref{eq.25}):

 \begin{equation}
 \Delta \theta_s^t = \theta_s^{t+1} - \theta^t 
 \label{eq.25}
 \end{equation}

 \item The server aggregates updates using the Federated Averaging algorithm as described in Eq.~(\ref{eq.26}):

 \begin{equation}
 \theta^{t+1} = \theta^t + \frac{1}{S} \sum_{s=1}^{S} \Delta \theta_s^t \label{eq.26}
 \end{equation}
\end{enumerate}

\section{Proposed Framework}\label{PF}
\begin{figure*}[!ht]
 \centering
 \includegraphics[width=\textwidth]{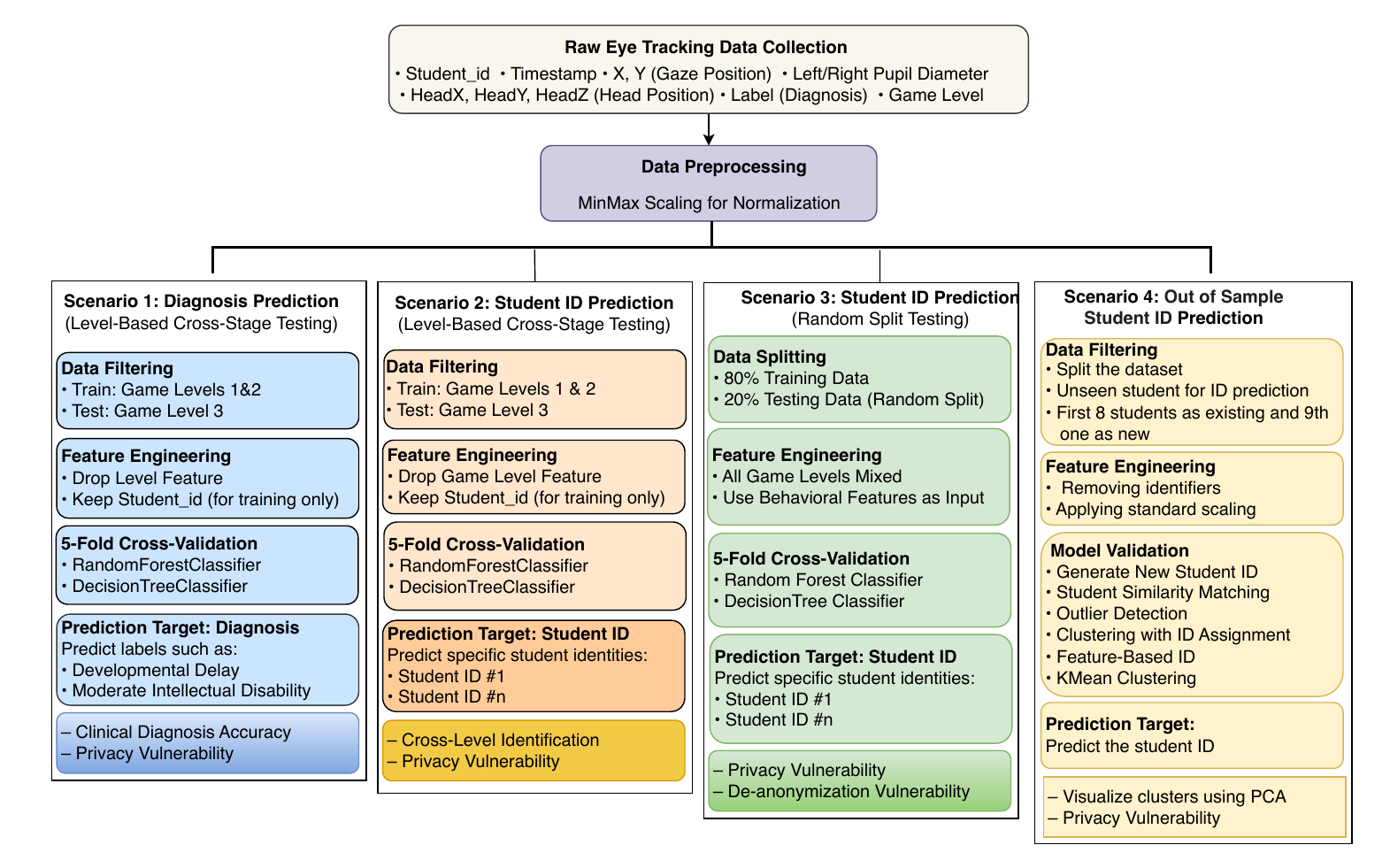}
 \caption{Eye Tracking Data-based Backtracking Scenarios}
 \label{fig3}
\end{figure*}

Fig. \ref{fig3} illustrates four scenarios of Phase 1 that utilize eye-tracking data for diagnosis identification and identity backtracking. Starting with the collection of raw data, this data is preprocessed through MinMax scaling for normalization before branching into four distinct scenarios. In Scenario 1 of diagnosis prediction, models are trained on game levels 1 and 2 and tested on level 3. Then, the game level feature is removed, and the student ID is retained solely for training purposes. The Models, including RF and DT classifiers, predict diagnosis labels such as Moderate Intellectual Disability, Developmental Delay, etc.) In Scenario 2 of student ID prediction, the same level-based split (training on levels 1-2 and testing on level 3) is used, but the target is changed to predict Student IDs instead of diagnoses. Both ML models predict specific student identities. This raises significant privacy implications, which indicate that biometric signatures persist across game levels. In Scenario 3 of student ID prediction using random split testing, the data is randomly split into training and testing sets, with all game levels combined and behavioral features used as input. The same ML models are then used to predict Student IDs. This likely yields higher accuracy but poses the most serious privacy risks and represents the most serious de-anonymisation vulnerability. All scenarios use 5-fold cross-validation with RF and DT classifiers except scenario 4. Still, their implications differ dramatically, ranging from legitimate medical use to privacy vulnerabilities that could enable tracking individuals across various contexts. A significant strength of this approach lies in its handling of unseen cases through a dedicated out-of-sample prediction module (Scenario 4). This includes a combination of K-Nearest Neighbours with confidence scoring, outlier detection via Isolation Forest, feature-based hash ID generation, and clustering with K-Means and PCA \footnote{\url{https://scikit-learn.org/stable/modules/generated/sklearn.decomposition.PCA.html}} visualisation. This hybrid strategy enables the framework to identify known individuals with high accuracy and to flag and assign IDs to new students who were not present in the training data.

\subsection{Serious Game Model, Dataset, and Preliminaries}
% \textcolor{red}{==========ARE=================}
% \textcolor{red}{WORK IN PROGRESS}

\subsubsection{Game structure}
\begin{figure}[!ht]
 \centering
 \includegraphics[width=0.8\linewidth, angle=-90]{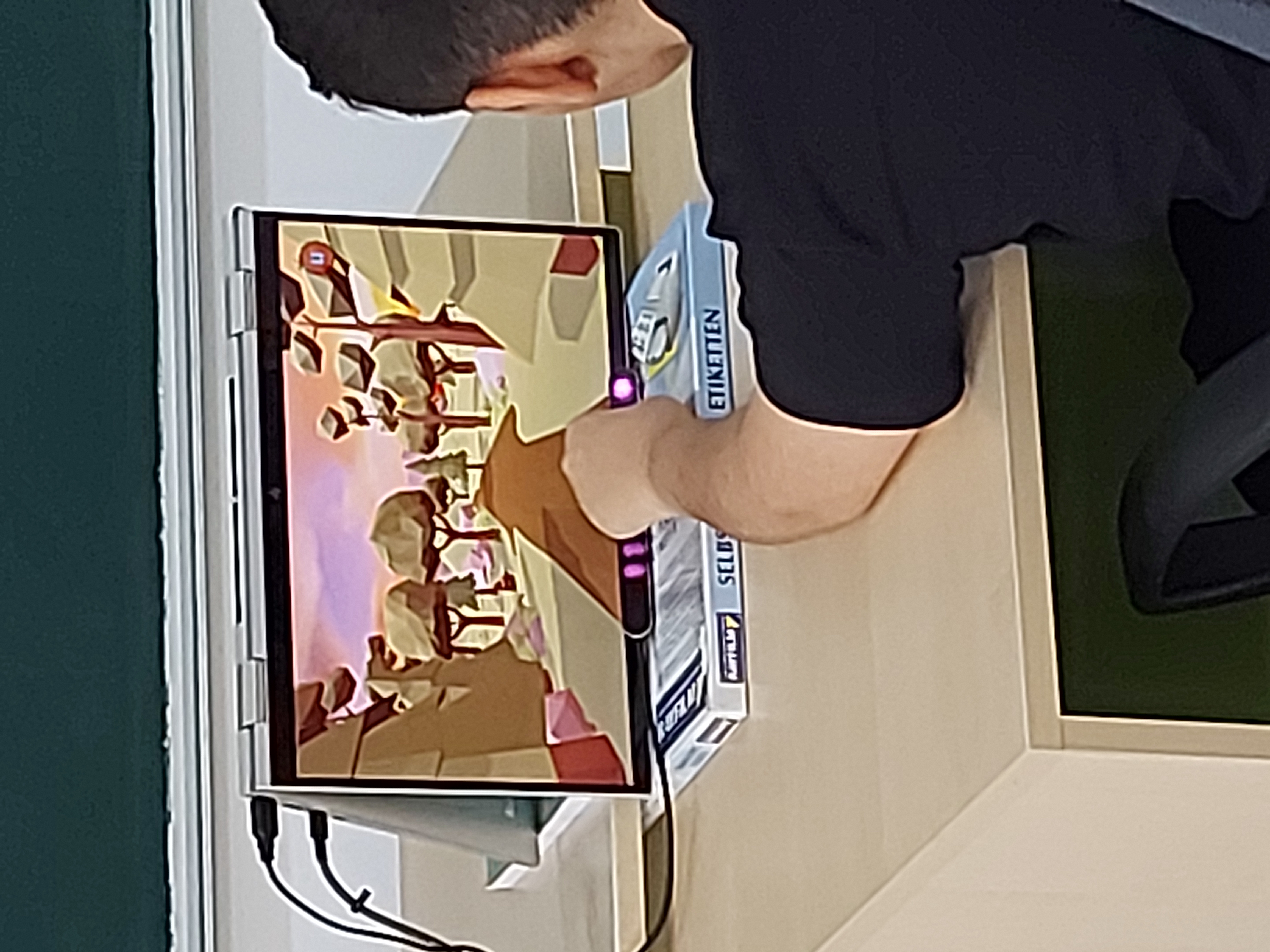}
 \caption{A participant performing the CCPT task while eye-tracking data is being collected}
 \label{figParticipantPhoto}
\end{figure}

The sustained attention game used for data collection was formally structured and designed by Costescu et al.~\cite{OriginalMushroomGamePaper}. It follows the original task design developed by Rosvold et al.~\cite{CCPT} as an implementation of the Computerized Continuous Performance Task (CCPT). Participants are shown a road in the forest, where mushrooms (\textit{targets}) and flowers(\textit{distractors}) appear randomly on either side. The objective is to touch the screen when a target appears while avoiding interaction with distractors. A Tobii Pro Nano eye-tracker was used for data collection, and the game utilizes a custom calibration and head positioning screen explicitly designed for the target demographic. Fig. \ref{figParticipantPhoto} shows a participant playing the sustained attention game while gaze data is collected. 

\subsubsection{Data Collection Protocol}
\begin{figure}[!ht]
 \centering
 \includegraphics[width=0.8\columnwidth]{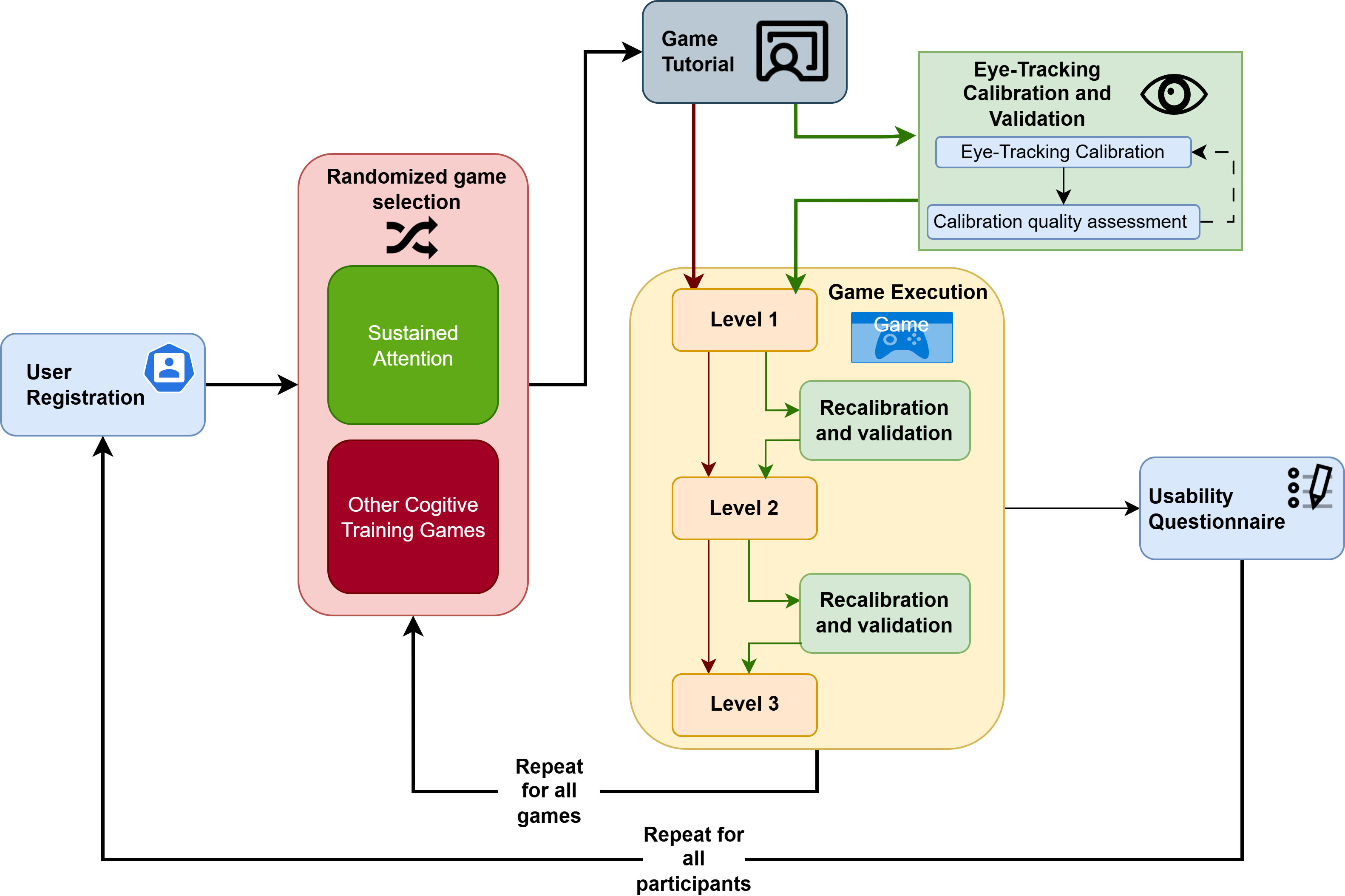}
 \caption{Data collection protocol}
 \label{figDataCollection}
\end{figure}

The participant pool for data collection consisted of 12 school-age participants with NDD, aged 9 to 14. Data was collected at a special education school in Romania. Each data collection session lasted approximately 1 hour and consisted of three games designed to train different cognitive functions, each with three levels. Our dataset is constructed from data collected for the EU-funded EMPOWER project (grant agreement No. 101060918, which includes 9 games designed to teach regulation strategies for various cognitive functions. While all games are necessary for the goals of the EMPOWER project~\cite{projectempowerProjectEMPOWER}, eye-tracking data collection features were a priority in all games at this project stage. Gaze data were not collected during most games to reduce session length from calibration procedures and maintain a more reliable dataset. Only gaze data from the sustained attention game was collected to construct our dataset. Fig. \ref{figDataCollection} illustrates the data collection protocol.

As data collection was performed on-site at special education schools in different countries as part of the EMPOWER project, the necessity for a mobile setup and varying environmental conditions directly impacted the quality and reliability of our eye-tracking dataset. This is further complicated by the placement of this game within the collection procedure, which was randomized to reduce order bias. This introduces a vulnerability to different fatigue levels during data collection, and both vulnerabilities can directly affect the quality of eye-tracking data~\cite{ETDataQuality}. To mitigate these issues in our dataset, the following measures were implemented:

\begin{itemize}
 \item Eye-tracking operators were available on-site during all data collection events.
 \item Strict optimal setup requirements were constructed to ensure the best possible data quality in various settings.
 \item The calibration procedure was optimized and adjusted to maintain the engagement of participants within the desired age range~\cite{CalibrationEyeTrackingChildren}. This includes custom graphics for fixation points and the addition of audio cues.
 \item Additional data pre-processing was done to ensure the participants' validity in the final dataset.
\end{itemize}

After excluding 3 participants due to data quality issues, the models in the proposed framework are trained on gaze data collected from 9 students.

\begin{algorithm}[!ht]
\caption{Proposed Pseudo-code for Diagnosis and Student ID Prediction}
\label{Pseudocode}
\begin{algorithmic}[1]
\STATE \textbf{Input:} Raw dataset $(D)$ with gaze, pupil dilation, head movement (x, y), labels, threshold
\STATE \textbf{Output:} Predicted diagnosis or student ID (known or new)
\STATE \textbf{Pre-processing}
 \STATE \quad Clean missing/duplicate entries
 \STATE \quad  Encode categorical labels (diagnosis, game level, student ID)
 \STATE \quad  Normalize Dataset
\STATE \textbf{Diagnosis Prediction}
 \STATE  \quad Train Random Forest and Decision Tree Classifiers
 \STATE \quad  Utilize 5-fold cross-validation
 \STATE \quad  Output: Diagnosis label
\STATE \textbf{Student Classification (In-Sample)}
 \STATE \quad  Train models using known student data
 \STATE \quad  Evaluate Performance Metrics

\STATE \textbf{Student Prediction (60-40 Split)}
 \STATE \quad  Train models on 60\% of known data
 \STATE \quad  Test on 40\% of the unseen data
 \STATE \quad  Evaluate generalization performance

\STATE \textbf{Out-of-Sample Student ID Prediction}
 \STATE  \quad Standardize new input data using a saved scaler
 \STATE \quad  Predict nearest known student using KNN ($k=1$)
 \STATE \quad  Calculate confidence: $\text{confidence} = \exp(-\text{distance})$
 \IF {confidence $<$ threshold}
 \STATE Mark as a new student
 \ELSE
 \STATE Assign ID of the nearest student
 \ENDIF
 \STATE Use Isolation Forest to detect outliers (new profiles)
 \STATE Generate a hashed ID from key features if no match is found
 \STATE Apply K-Means clustering to check if the student is in a known cluster
 \STATE Visualize using PCA if needed
\RETURN Predicted label or new ID assignment
\end{algorithmic}
\end{algorithm}

\subsection{Feature Extraction}
In this study, the raw input data comprises multidimensional time-series streams captured during gameplay or task interaction, including eye-tracking data (gaze coordinates), physiological signals (pupil dilation), and head movement information in three axes (HeadX, HeadY, HeadZ). Each modality carries informative behavioral patterns that can indicate cognitive profiles and individual identities. To transform this raw data into usable features, we begin by aggregating the values across sessions, ensuring temporal alignment and consistency. Gaze coordinates are treated as continuous variables capturing spatial attention. At the same time, left and right pupil dilation values are averaged and smoothed to mitigate the effect of short-term noise or camera jitter. After normalization, the head movement dimensions are included as-is to retain their full dynamic range, which helps capture motor behavior. In addition to raw measurements, derived features such as variance, range, and differential movement can be integrated into future work to gain a more profound insight.
Categorical variables, such as the diagnostic label and game level, are transformed using one-hot or label encoding depending on the target task. Specifically, for classification tasks involving diagnosis, the label is encoded as an integer class index, while the student ID is treated as a unique class for multiclass prediction. To prepare the data for machine learning, all numerical features are standardised using z-score normalisation with StandardScaler, resulting in a zero mean and unit variance for each feature. This step is crucial because it ensures that features with different scales (such as pixel distances for gaze and decimal values for dilation) do not disproportionately influence model training. The final feature set thus consists of a uniform vector representation for each session, ready for ingestion into classifiers and clustering algorithms.

\subsection{Cross-validation Training}
To ensure that the models built for diagnosis classification and student identification are generalizable and not overfitted to the training data, we employ k-fold cross-validation as a core evaluation strategy. In our experiments, a 5-fold cross-validation scheme is employed, which involves splitting the dataset into five equally sized parts. In each round of validation, four folds are used for training, and the rest are used for testing. The final performance metrics are averaged across all five folds to estimate the model's generalisation performance robustly. This method helps mitigate issues of variance and bias that might arise from a single train-test split and is particularly important when dealing with limited data or class imbalance, as is common in educational or medical datasets.
For multiclass classification tasks such as student ID prediction, stratified k-fold is implicitly applied to maintain a proportional representation of each class in every fold. In models such as the Random Forest Classifier and Decision Tree Classifier, cross-validation also enables model selection by tuning hyperparameters based on the stability of performance across multiple folds.
\subsection{Phase 1 Setting}
\subsubsection{Scenario 1}
For the task of disorder diagnosis, we implemented a targeted training strategy where data from Levels 1 and 2 were used to train the models, and evaluation was conducted exclusively on Level 3. This approach assessed the models' ability to generalize across cognitive task difficulties. Two supervised learning algorithms were employed: the Random Forest \footnote{\url{https://scikit-learn.org/stable/modules/generated/sklearn.ensemble.RandomForestClassifier.html}} and the Decision Tree \footnote{\url{https://scikit-learn.org/stable/modules/generated/sklearn.tree.DecisionTreeClassifier.html}} Classifier. The Random Forest model was selected for its ensemble-based structure and robustness to noise and overfitting. It was configured with (n\_estimators=100, max\_depth=None, random\_state=42). In parallel, the Decision Tree Classifier was chosen for its interpretability and clear decision boundaries. It utilized the Gini impurity criterion for node splitting and maintained consistency across experiments by using the same random state setting.
Training was performed on the subset of data corresponding to Levels 1 and 2, ensuring that the models learned diagnostic patterns from earlier stages of the data. Testing was then carried out using data from Level 3, providing a realistic scenario where the framework is evaluated on more challenging tasks. We employed 5-fold cross-validation within the training set to ensure reliable performance measurement and assessed final performance on the held-out Level 3 data. The evaluation included confusion matrices and classification reports to analyse model behavior across diagnostic classes. 

\subsubsection{Scenario 2}
To evaluate the model's ability to generalize across task complexity, we trained our student identification models on data from Levels 1 and 2 and tested them exclusively on Level 3 data. In this setup, each student is treated as a separate class, and the objective is to accurately predict student identities based solely on their raw gaze data as they progress through increasingly challenging gameplay. The models used include Random Forest and Decision Tree classifiers, with hyperparameters consistent with earlier experiments: 100 estimators for Random Forest, no depth restriction, and the Gini index as the splitting criterion for Decision Trees. This experimental configuration enables us to evaluate how effectively the framework captures stable, individual-specific gaze patterns that persist even as task difficulty increases.

\subsubsection{Scenario 3}
To evaluate the model's ability to generalize to partially unseen data, we perform a randomised split of the dataset into two partitions: 60\% for training and 40\% for testing. This division reflects a practical scenario where the model must learn from a limited subset of students and then predict students' identities from a hold-out set. Using their standard configurations, the training set is used to fit the Random Forest and Decision Tree classifiers. Specifically, the Random Forest model employs 100 estimators without depth restriction and a random\_state=42 to ensure reproducibility. In contrast, the Decision Tree model uses the Gini criterion for splitting and a similar random seed. The testing set is then used to assess the model's predictive accuracy, analyse confusion matrices, and compute per-class performance metrics. This 60/40 setup provides a controlled environment to measure the robustness of the student identification framework and establishes a foundational benchmark for comparison with out-of-sample prediction methods.

\subsubsection{Scenario 4}
\begin{figure}[!ht]
 \centering
 \includegraphics[width=0.8\columnwidth]{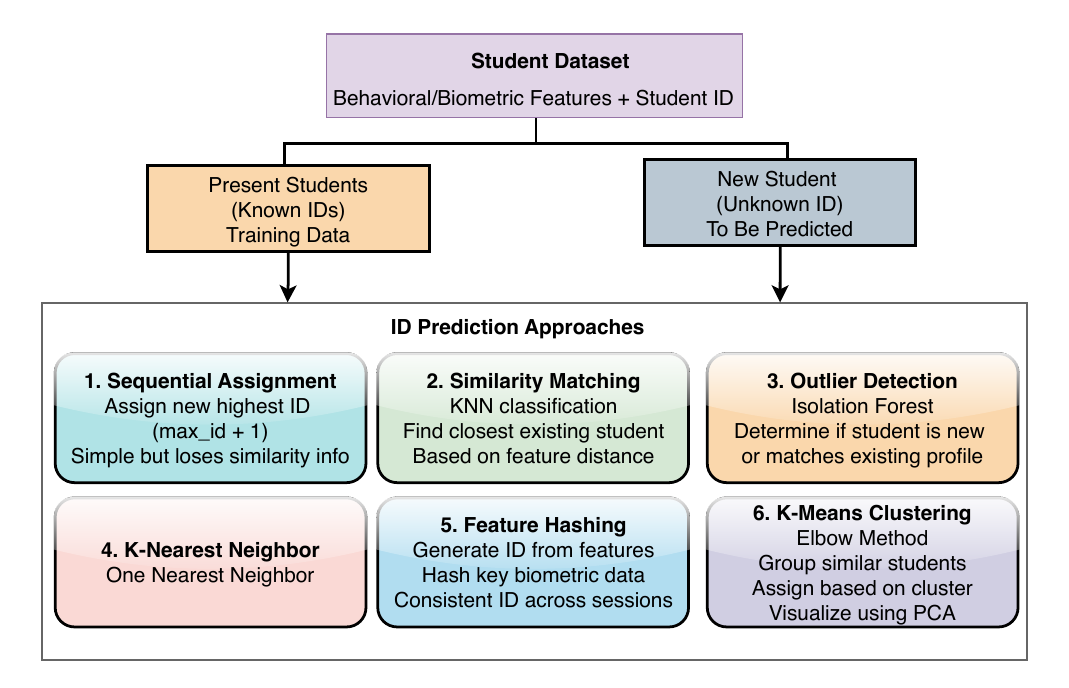}
 \caption{AI Approaches for Student ID Prediction/Assignment}
 \label{fig4}
\end{figure}

We introduce a robust, multi-stage modelling strategy centred around K-means clustering to address the challenge of identifying student profiles that have not been previously encountered during training. In this scenario, the first eight students are treated as known (in-sample), while the ninth is considered entirely new (out-of-sample). The primary goal is to determine whether the ninth student can be matched to existing profiles or should be assigned a new identifier.
K-Means clustering is an unsupervised method for segmenting known students into distinct behavioral clusters. The optimal number of clusters is estimated using the elbow method, typically ranging from 2 to 6 clusters, with random\_state=42 and n\_init=10 for consistency and convergence reliability. Once the model is trained on the standardised feature vectors of the first eight students, the ninth student's data is also scaled and projected into the same feature space. The framework calculates the Euclidean distance from the new student's feature vector to each cluster centre, and the shortest of these distances is considered the student's proximity to known behavioral patterns.
To determine whether the new student falls within the boundaries of existing clusters, we define a dynamic threshold based on the 95th percentile of the minimum distances from the existing students to their nearest cluster centers. Suppose the ninth student's distance to its nearest cluster exceeds this threshold. In that case, the framework classifies them as an outlier, implying that their behavior deviates significantly from the known population and warrants a new student ID assignment.
To aid interpretability, Principal Component Analysis (PCA) is used to reduce the high-dimensional feature space to two principal components, enabling visual inspection of how closely the new student aligns with existing clusters. This step supports intuitive validation, highlighting cluster separation and overlap. In essence, this K-Means-based framework forms a crucial decision-making layer in our out-of-sample prediction module, combining clustering distance, percentile-based thresholds, and dimensionality reduction to handle new students in a scalable, explainable manner.
Fig. \ref{fig4} depicts the key approaches to student ID prediction and assignment methods. It starts with a dataset that contains behavioral and biometric features, which is split between existing students (with known IDs) and new students (whose IDs need to be predicted). Six main approaches are used for ID prediction.
\begin{enumerate}
\item Sequential Assignment: Assign the new highest ID (max\_id + 1), which is simple but discards similarity information.

\item Similarity Matching: Uses KNN classification to find the closest existing student based on feature distance.

\item Outlier Detection: Employs Isolation Forest to determine if a student is new or matches an existing profile.

\item Clustering: Applies K-Means to group similar students and make assignment decisions based on the cluster.

\item Feature Hashing: Generates IDs directly from the biometric features, creating consistent session identifiers.

\item Ensemble framework: Combines these multiple approaches with a confidence threshold to decide whether it is a new ID or a match to an existing ID.
\end{enumerate}
High accuracy in ID prediction indicates the persistence of biometric signatures in eye tracking data, which has significant privacy implications.

\subsection{Phase 2 Setting}
This section covers privacy-preserving techniques for securing students' data during AI model training and protecting their original IDs.
\subsubsection{Dual-Layer Protection and Federated Learning Implementation}
\begin{figure}[!ht]
 \centering
 \includegraphics[width=0.8\linewidth]{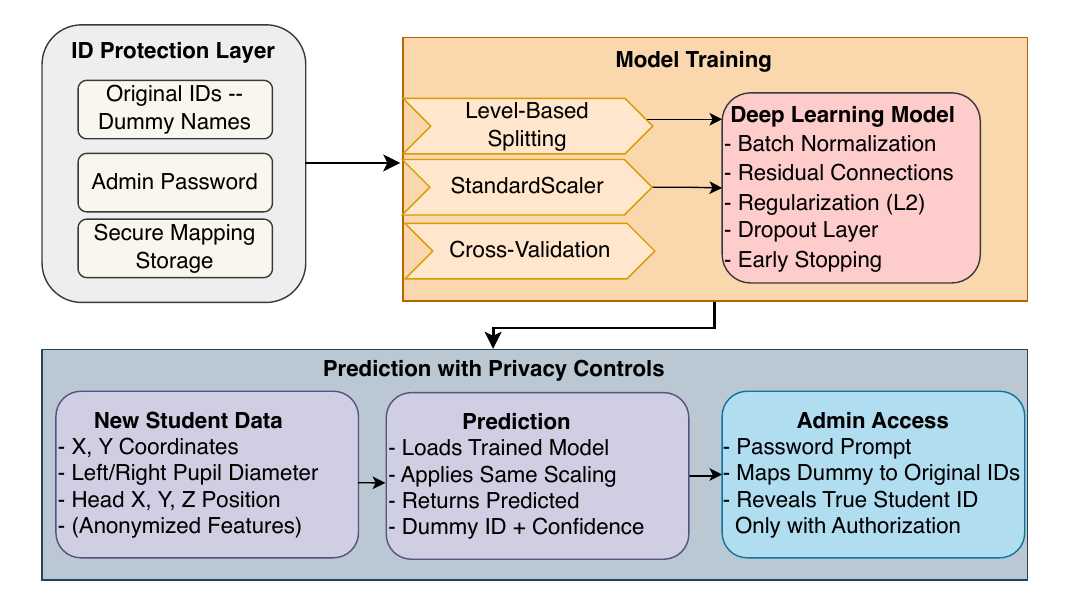}
 \caption{Dual-Layered approaches to Secure Student ID framework}
 \label{fig5}
\end{figure}

Fig. \ref{fig5} illustrates the multi-layered approaches for the secure student ID framework with privacy protection. First, an ID Protection Layer converts original student IDs into dummy names. These dummy names are securely mapped to the real IDs, which can only be accessed with an admin password. Then, the Model Training Pipeline follows a structured learning approach by splitting data based on game levels. StandardScaler is applied, and cross-validation is performed. The neural network is trained using advanced techniques, including batch normalization, residual connections, L2 regularization, dropout layers, and early stopping. For Prediction with Privacy Controls, when new student data, including gaze coordinates, pupil diameter, and head position, is input, the trained model applies the same scaling methods and generates a predicted dummy ID with a confidence score. Finally, Admin Access requires a password prompt to map the dummy ID back to the original student ID, ensuring that true identities are only revealed with proper authorization. This two-layer security approach combines dummy ID anonymisation and password access control, significantly securing student identification and reducing privacy risks.

We devised a dual-layer architecture that first generated dummy IDs against the original IDs. We also created an administration layer that can help map the dummy IDs to the original IDs, in case any stakeholder wants to view the original ID. FL's advanced distributed machine learning concept permits various clients, including mobile devices and edge servers, to train shared models through collaboration without publicizing their data. Traditional centralized learning involves the central aggregation of training data at a single location. However, FL allows clients to work locally with their information while sending model parameters, such as gradients or weights, to the central server. By dispersing computation across different users, the framework achieves enhanced data protection and operational efficiency, making it ideal for operations in the healthcare and finance sectors that handle confidential information. FL operates through successive training rounds, in which clients use local data to develop model updates. These updates are accumulated by the central aggregator through Federated Averaging protocols to advance the global model's effectiveness. 

\subsubsection{Client Model Architecture}
\begin{figure}[!ht]
 \centering
 \includegraphics[width=0.9\columnwidth]{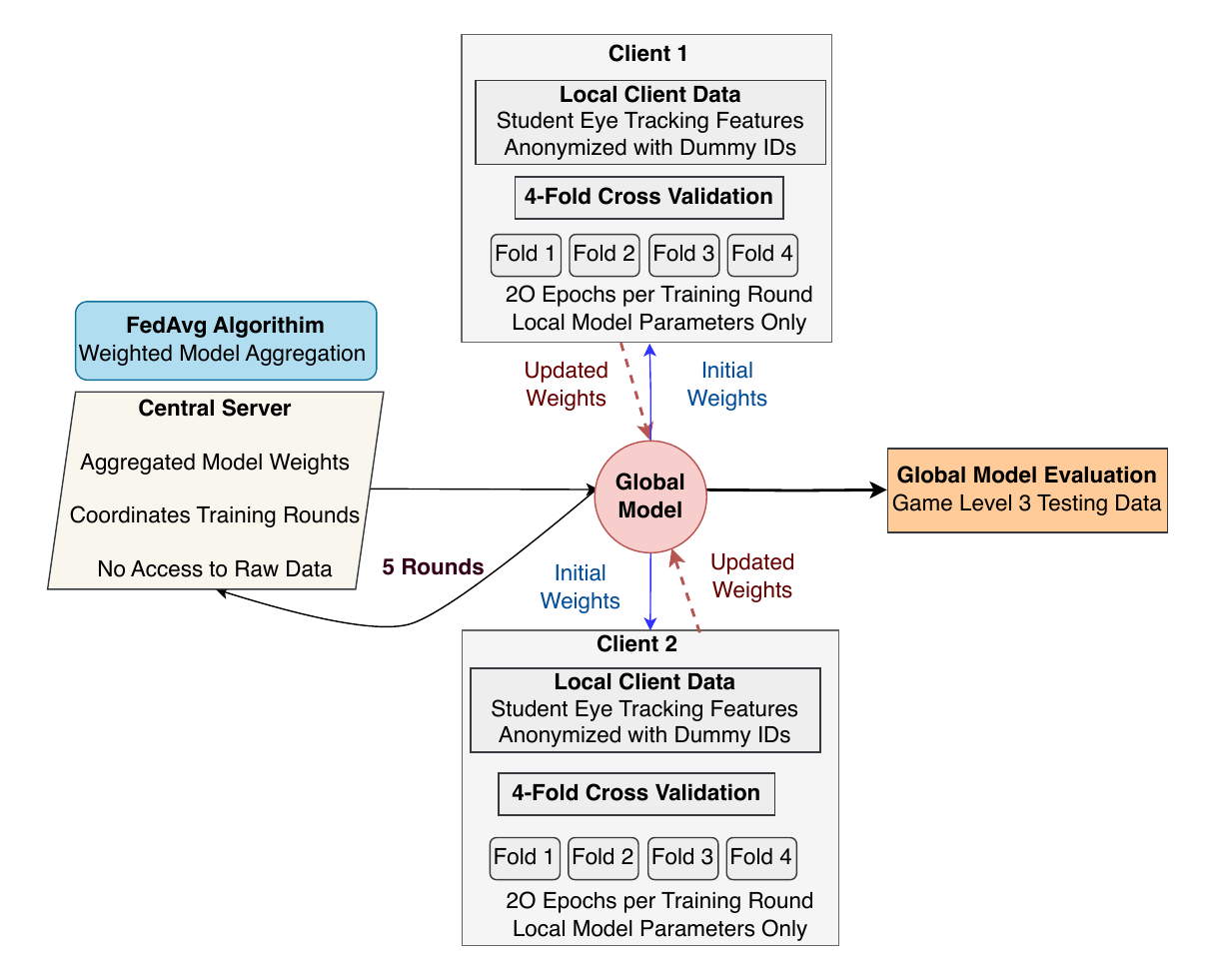}
 \caption{Secure Student ID prediction using FL process}
 \label{fig6}
\end{figure}

The client model architecture in FL serves as a fundamental element that shapes both the speed and efficiency of the learning process during operations. Fig. \ref{fig6} depicts the FL process for a secure student ID framework. The workflow has several key components: The central server coordinates the FL process, but never accesses raw student data. It maintains a global model that aggregates knowledge from both clients. Each client (Client 1 and Client 2) works with its local dataset of eye-tracking features using anonymized dummy IDs for student identification. Both clients perform independent 3-fold cross-validation training with 25 epochs per round.
During the FL process:
\begin{enumerate}
 \item The global model shares its weights with both clients.
 \item Each client trains locally on their private data.
\item Only model weights (not raw data) are returned to the central server.
\item The server aggregates these weights using the FedAvg algorithm.
\item This cycle repeats for 5 rounds to improve the global model.
\end{enumerate}

The framework includes two security layers: one is an ID protection layer that replaces real student IDs with dummy names, and the second is Password-protected administrator access to reveal the original student IDs. After training, the global model is evaluated on a separate test dataset (Game Level 3) to measure prediction accuracy.

Deep neural networks (DNNs) function as the model architecture that performs multi-class classification duties according to Algorithm \ref{alg}. The input layer accepts feature vectors to start the process, which is followed by dense layers that implement L2 regularisation (Kernel\_regulizer ($K_{R)}$)) for the prevention of overfitting. Combining Batch Normalisation (BN) across the entire network for normalisation, which overlaps with LeakyReLU (L\_R) activation functions, keeps neurons responsive to small negative input values. During training, the model applies dropout layers that randomly deactivate neurons to reduce dependence on specific features while enhancing generalisation capabilities.

\begin{algorithm}[!ht]
 \caption{Deep Neural Network Model Architecture and Parameters}
 \label{alg}
 \begin{algorithmic}[1]
 \STATE \textbf{Input:} Dataset, Number of classes
 \STATE \textbf{Output:} Trained Deep Neural Network
 \STATE Initialize input layer $inputs \leftarrow Input(input\_shape)$
 \COMMENT{First block with residual connection}
 \STATE $x \leftarrow BN(inputs)$
 \STATE $x \leftarrow Dense(256, K_{R}=l2(0.001))(x)$
 \STATE $x \leftarrow BN(x)$
 \STATE $x \leftarrow L\_R(\alpha=0.1)(x)$
 \STATE $x \leftarrow Dropout(0.3)(x)$

 \COMMENT{Intermediate layer}
 \STATE $x \leftarrow Dense(128, K_{R}=l2(0.001))(x)$
 \STATE $x \leftarrow BN(x)$
 \STATE $x \leftarrow L\_R(\alpha=0.1)(x)$
 \STATE $x \leftarrow Dropout(0.2)(x)$

 \COMMENT{Residual block}
 \STATE $skip \leftarrow x$
 \STATE $x \leftarrow Dense(128, K_{R}=l2(0.001))(x)$
 \STATE $x \leftarrow BN(x)$
 \STATE $x \leftarrow L\_R(\alpha=0.1)(x)$
 \STATE $x \leftarrow Dropout(0.2)(x)$
 \STATE $x \leftarrow Dense(128, K_{R}=l2(0.001))(x)$
 \STATE $x \leftarrow BN(x)$
 \STATE $x \leftarrow L\_R(\alpha=0.1)(x)$
 \STATE $x \leftarrow x + skip$
 \STATE $x \leftarrow Dropout(0.2)(x)$

 \COMMENT{Output layer}
 \STATE outputs $\leftarrow$ Dense(num\_classes, activation= 'softmax')(x) 
 \STATE $model \leftarrow Model(inputs, outputs)$
 \STATE Compile model with Adam optimiser and sparse categorical cross-entropy loss \COMMENT{Create and compile model}
 \RETURN $model$
 \end{algorithmic}
\end{algorithm}

This model demonstrates strength from its residual block, an essential architectural element within deep architectures, including ResNet.
During backpropagation, residual connections facilitate a smooth gradient flow, resolving problems such as vanishing gradients and enhancing the overall training procedure. The network terminates with a softmax output layer, making it suitable for classifying problems that involve multiple categories. The model utilises the Adam optimiser as one of the standard adaptive learning rate algorithms, together with a sparse categorical cross-entropy loss that handles integer-encoded multi-class labels effectively. The capacity of this model to function successfully in an FL environment directly corresponds to its ability to maintain performance across various decentralised data distributions. The visualisation shows how training data is distributed across clients (Client 1 and Client 2) by displaying student ID categories. The bar chart indicates that client data show equivalent distributions for student populations, leading to balanced learning without significant heterogeneity effects in the data. Model convergence benefits from this distribution since clients hold training samples with similar numbers of data points. Real-world FL applications often encounter data heterogeneity issues, necessitating specialized approaches such as clustered client scheduling or personalized FL to standardize client performance outcomes.

\subsubsection{Cross-validation Strategy}
Stratified K-Fold Cross-Validation is vital in building robust, generalizable client models. The decentralised nature of FL produces client datasets of various sizes and class distribution levels. A strategic training method requires cross-validation to address the challenges of data imbalance and scarcity resulting from data variability. StratifiedKFold ensures proper distribution of different classes between folds to achieve balanced learning by maintaining class proportions in each split.

Algorithm \ref{alg2} illustrates the training process, which begins by dividing the client dataset into n\_folds using the default value of NUM\_FOLDS, while retaining the original class distribution in each subset.
The stratified data division method produces optimal results when analysing imbalanced datasets because it blocks models from developing biased preferences towards major groups. A new model instance is created within each fold and initializes itself with the global model weights (which must be available and compatible). The initialisation procedure enables clients to leverage previously accumulated knowledge from their model while adapting their framework to handle new local data. If structural differences prevent the weights from being usable, the function automatically switches to conducting new training without prior weights. The training process contains two optimisation techniques: early stopping ($E_{s}$) and learning $L_{r}$ rate scheduling for enhancement. $E_{s}$ stops training as validation loss becomes stagnant, while ReduceLROnPlateau automatically controls the learning rate to speed up convergence. The combined features stop modelling from becoming excessive and maintain adequate learning progress. After processing, all folds of the model with the highest peak validation accuracy are selected as the final model.

\begin{algorithm}[!ht]
\caption{FL with K-Fold Cross-Validation}
\label{alg2}
\begin{algorithmic}[1]
\REQUIRE Client data $D = \{X, y\}$, global weights $W_g$ (optional), input shape $S$, number of folds $n\_folds$, epochs $E$
\ENSURE Best trained model $M_{best}$, best validation accuracy $Valid_{best}$

\STATE Initialise Stratified K-Fold with $n\_f$
\STATE $Models \gets []$, $Accuracies \gets []$

\FOR{$fold = 1$ to $n\_folds$}
 \STATE Split $D$ into $(X_{train}, y_{train})$ and $(X_{val}, y_{val})$ for each $fold$
 \STATE Initialize new model $M$ with input shape $S$
 
 \IF{$W_g$ is available}
 \STATE Load $W_g$ into $M$ if available
 \IF{Loading fails}
 \STATE Train from scratch
 \ENDIF
 \ENDIF
 
 \STATE Define callbacks:
 \STATE $\quad E_{s}$ (patience = 10, restore best weights)
 \STATE $\quad$ ReduceLROnPlateau (patience = 5, factor = 0.5, min lr = $10^{-6}$)

 \STATE Train $M$ using $(X_{train}, y_{train})$ for $E$ epochs with batch size = 64 and validate on $(X_{val}, y_{val})$
 
 \STATE Compute validation accuracy $Valid_{acc}$
 \STATE Append $M$ to $Models$ and $Valid_{acc}$ to $Accuracies$

\ENDFOR

\STATE Find index $best\_idx$ where $Accuracies$ is maximum
\STATE $M_{best} \gets Models[best\_idx]$
\STATE $Valid_{best} \gets Accuracies[best\_idx]$

\RETURN $M_{best}, Valid_{best}$
\end{algorithmic}
\end{algorithm}

\subsubsection{Federated Averaging Algorithm}
In the aggregate\_models() function, Federated Averaging (FedAvg) \cite{sun2022decentralized} combines model updates sent by multiple clients to produce one unified global model. The function starts by verifying the availability of client weights derived from dataset sizes. The absence of client weights leads to equal treatment of all clients during aggregation. The framework executes normalization on weights before summation to achieve a total value of 1 for accurate weight average calculations. As a first step, the function creates an empty list that matches the weight structure of the first client model, but all values are set to zero. During the iteration, the function retrieves all client model weights sequentially, applying the normalized weights for weighted computation. Clients with larger datasets have a greater impact on the resulting model through this method. After aggregating all client contributions, the function produces an updated version of global model weights for the subsequent federated training round. The framework adheres to FedAvg principles by allowing decentralized data processing on local models before distributing only weights between nodes to maintain privacy. FedAvg establishes client model averaging based on dataset proportions to enhance model stability while enabling equitable participation from all clients, making it a crucial method for FL.

\subsubsection{Privacy Vulnerability Assessment}
\begin{figure}[!ht]
 \centering
 \includegraphics[width=0.8\linewidth]{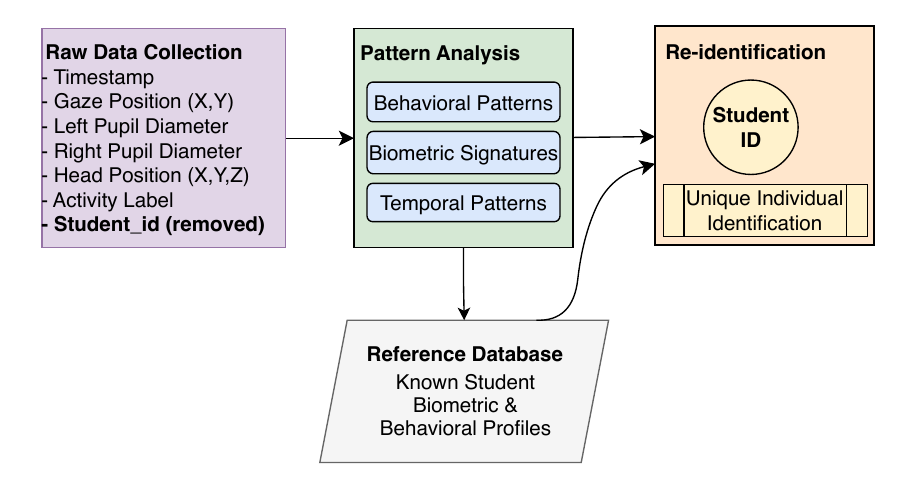}
 \caption{Backtracking Student IDs from Biometric features}
 \label{fig7}
\end{figure}

Fig. \ref{fig7} illustrates how a student's identity could be backtracked from the biometric and behavioral features collected through eye tracking, even when the Student\_ID is removed from the dataset. The process starts by collecting raw data (eye coordinates, pupil diameters, head position, timestamps, gaze positions, and activity labels), and then removes the student ID. Then, move on to pattern analysis, where machine learning algorithms can identify unique behavioral and temporal patterns, as well as biometric signatures. Finally, backtracking can be reached by matching these patterns against reference databases or known profiles of student biometrics and behavior. This figure highlights the privacy vulnerability in such datasets, even when student IDs are removed. Combining these biometric features can create a unique "fingerprint" that could be used to identify a specific individual. 

\section{Results and Analysis}\label{RA}
\subsection{Phase 1 Results}
\subsubsection{Scenario 1}
In this study, we evaluated the ability of classification models to generalize diagnostic predictions across varying cognitive contexts by training on data from game levels 1 and 2 and testing on the more cognitively demanding level 3. Despite the increased complexity and pronounced class imbalance in the training and testing sets, the models demonstrated strong generalization performance, as shown in Table~\ref{Table1and2}. The Random Forest classifier achieved a test accuracy of 98.74\%, while the Decision Tree slightly outperformed it with 99.3\% test accuracy. 

\begin{table}[!ht]
\centering
\caption{Performance of Random Forest (RF) and Decision Tree (DT) on out-of-level data (Level 3). Key: Moderate Intellectual Disability-MDI, Developmental Delay-DD}
\label{Table1and2}
\begin{tabular}{|l|l|c|c|c|}
\hline
\textbf{Model} & \textbf{Class} & \textbf{Precision} & \textbf{Recall} & \textbf{F1-Score} \\ \hline
\multirow{3}{*}{RF} & MDI & 0.99 & 1.00 & 0.99 \\ 
 & DD & 0.99 & 0.87 & 0.92 \\
 & Accuracy & & & 0.987 \\ \hline
\multirow{3}{*}{DT} & MDI & 0.99 & 1.00 & 1.00 \\
 & DD & 1.00 & 0.92 & 0.96 \\
 &  Accuracy & & & 0.993 \\ \hline
\end{tabular}
\end{table}

\begin{figure}[!ht]
\centering
\subfloat[Random Forest\label{Fig.2(a)}]{
\includegraphics[width=0.495\columnwidth]{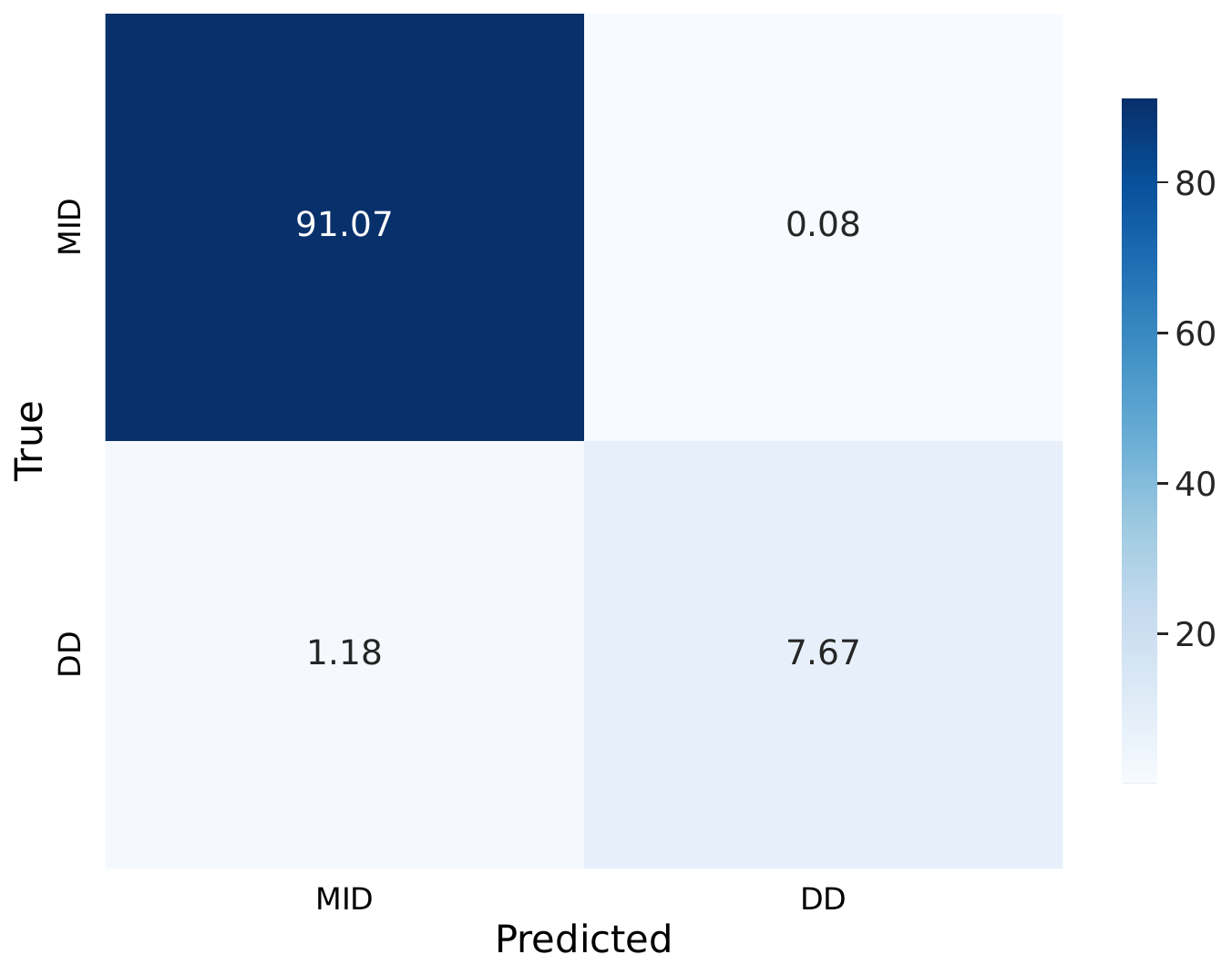}}
\subfloat[Decision Tree\label{Fig.2(b)}]{
\includegraphics[width=0.495\columnwidth]{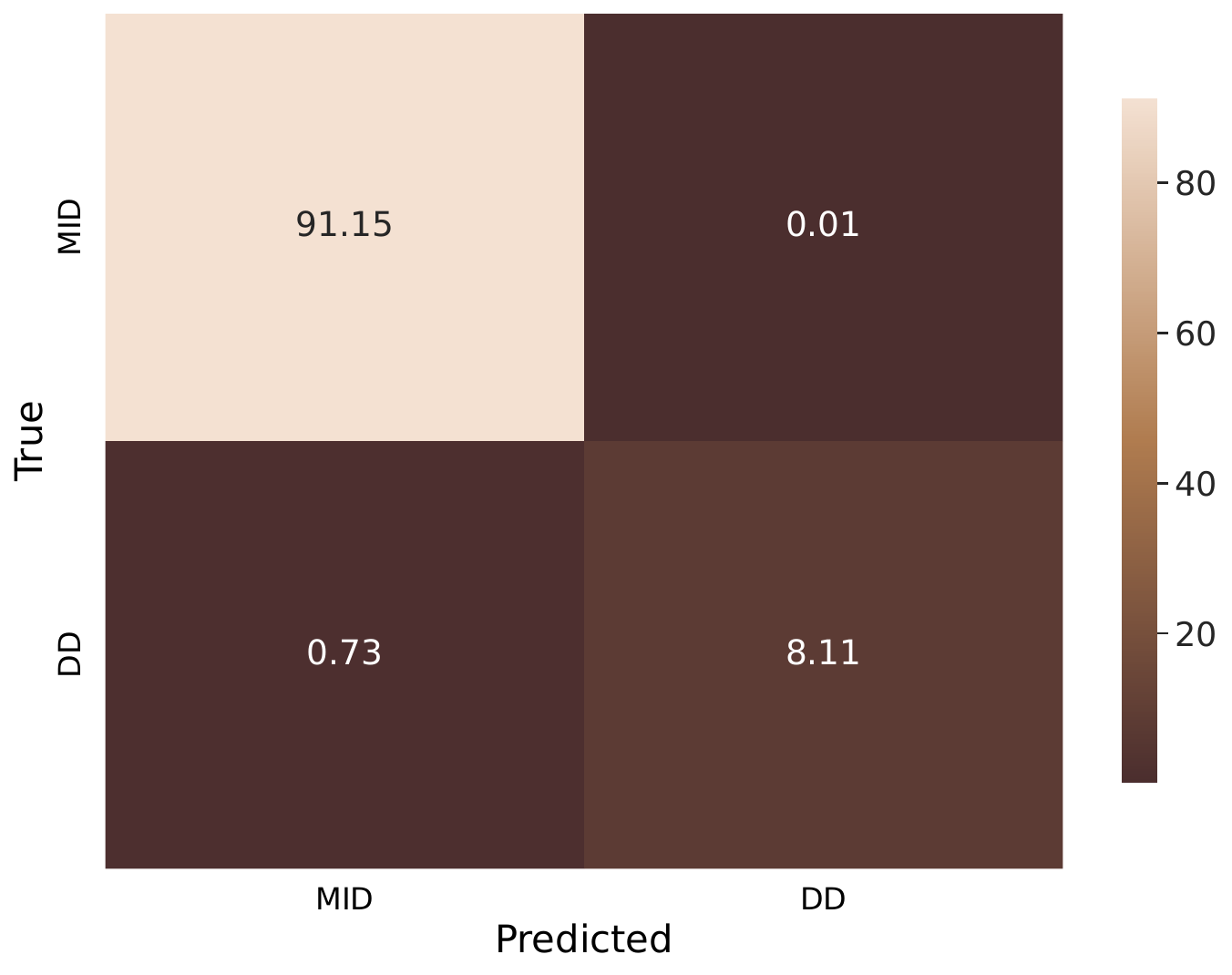}}\\
\subfloat[Feature Importance using Random Forest\label{Fig.2(c)}]{
\includegraphics[width=0.8\columnwidth]{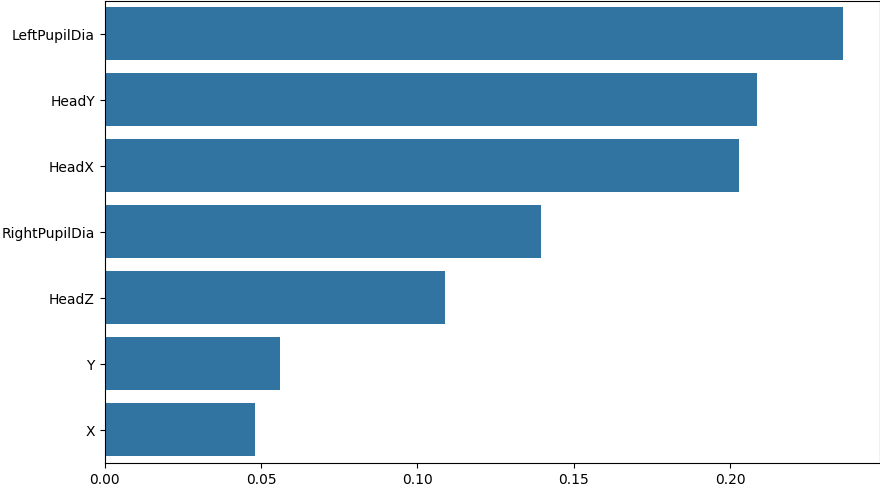}}
\caption{Performance evaluation of classification models under cross-context generalization: (a) Confusion matrix for the Random Forest classifier highlighting a significant misclassification trend for the minority class \textit{DD}, indicating susceptibility to class imbalance despite overall high accuracy; (b) Confusion matrix for the Decision Tree classifier showing improved minority class recall and a more balanced classification outcome compared to Random Forest; and (c) Feature importance ranking derived from the Random Forest model, emphasizing the predictive significance of physiological attributes such as \textit{RightPupilDia}, \textit{LeftPupilDia}, and head orientation features (\textit{HeadX}, \textit{HeadY}, \textit{HeadZ}).}

\label{Fig2}
\end{figure}

The confusion matrix for the Random Forest in Fig.~\ref{Fig2}(a) showed that among actual "DD" cases, 543 were correctly classified, and 3,538 were misclassified as "MDI." The model achieved high precision for actual "MDI" cases, though the specific values were not fully visible. The Decision Tree confusion matrix in Fig.~\ref{Fig2}(b) offered more complete information; it correctly classified 3,742 instances of "MDI" and 339 instances of "DD," while misclassifying 522 cases from each class, respectively, indicating a slightly better balance than the Random Forest. The Random Forest's feature importance plot in Fig.~\ref{Fig2}(c) identified RightPupilDia, LeftPupilDia, and head position features (HeadY, HeadX, HeadZ) as the most influential predictors, highlighting the relevance of physiological and behavioral cues in the classification task. These results collectively suggest that the models can effectively learn diagnosis-relevant patterns that transfer across levels of task difficulty, but also expose their sensitivity to class imbalance.

\subsubsection{Scenario 2}

We evaluated the ability of Random Forest and Decision Tree models to predict student identities based on raw gaze data, with training performed on game levels 1 and 2 and testing on the cognitively demanding level 3. The Random Forest achieved a 5-fold cross-validation accuracy of 91.93\%, but its accuracy dropped to 63.03\% on the testing set, indicating challenges in generalizing across varying task difficulties. The Decision Tree followed a similar trend with 89.05\% cross-validation and 59.91\% test accuracy. The classification reports in Table~\ref{Table3} revealed that students, such as Student 6 and Student 7, were accurately predicted, achieving F1-scores of 0.88 and 0.97, respectively, using the Random Forest. In contrast, others, like Student 1, were nearly unidentifiable (F1-score of 0.01) due to highly overlapping or underrepresented gaze patterns. 

\begin{table}[!ht]
\centering
\caption{Random Forest and Decision Tree classification report for student ID prediction}
\label{Table3}
\begin{tabular}{llccc}
\toprule
\textbf{Model}&\textbf{Student ID} & \textbf{Precision} & \textbf{Recall} & \textbf{F1-score} \\
\midrule
\multirow{9}{*}{RF}&1 & 0.01 & 0.01 & 0.01 \\
&2 & 0.00 & 0.00 & 0.00 \\
&3 & 0.78 & 0.47 & 0.58 \\
&4 & 0.51 & 0.99 & 0.67 \\
&5 & 0.15 & 0.03 & 0.05 \\
&6 & 0.80 & 1.00 & 0.88 \\
&7 & 0.99 & 0.95 & 0.97 \\
&8 & 0.86 & 0.62 & 0.72 \\
&9 & 0.84 & 0.88 & 0.86 \\
\midrule
\textbf{Accuracy} & && & \textbf{0.63} \\
\midrule
\multirow{9}{*}{DT}&1 & 0.00 & 0.00 & 0.00 \\
&2 & 0.00 & 0.00 & 0.00 \\
&3 & 0.60 & 0.47 & 0.53 \\
&4 & 0.64 & 0.99 & 0.77 \\
&5 & 0.43 & 0.21 & 0.28 \\
&6 & 0.79 & 0.99 & 0.88 \\
&7 & 0.95 & 0.43 & 0.59 \\
&8 & 1.00 & 0.66 & 0.80 \\
&9 & 0.45 & 0.83 & 0.58 \\
\midrule
\textbf{Accuracy} && & & \textbf{0.60} \\
\bottomrule
\end{tabular}
\end{table}

\begin{figure*}[!ht]
\centering
\subfloat[Random Forest Confusion Matrix\label{Fig.3(a)}]{
\includegraphics[width=0.45\textwidth]{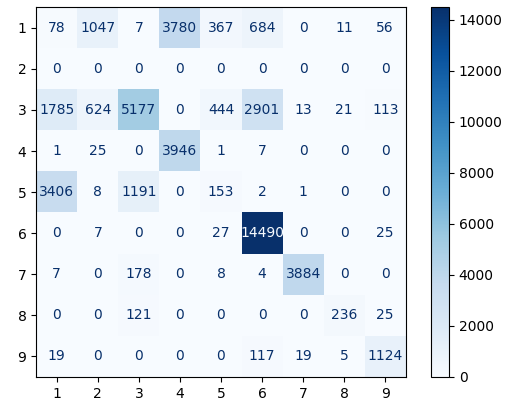}}
\subfloat[Decision Tree Confusion Matrix\label{Fig.3(b)}]{
\includegraphics[width=0.45\textwidth]{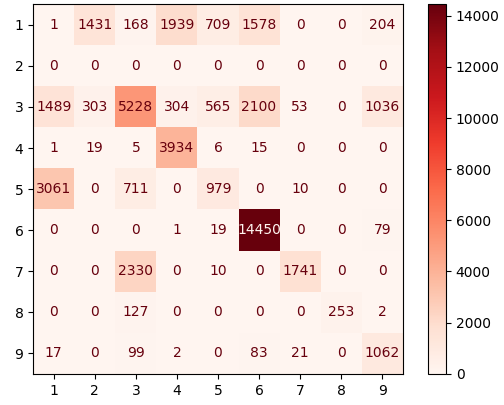}}\\
\subfloat[Feature Importance using Random Forest\label{Fig.3(c)}]{
\includegraphics[width=0.8\textwidth]{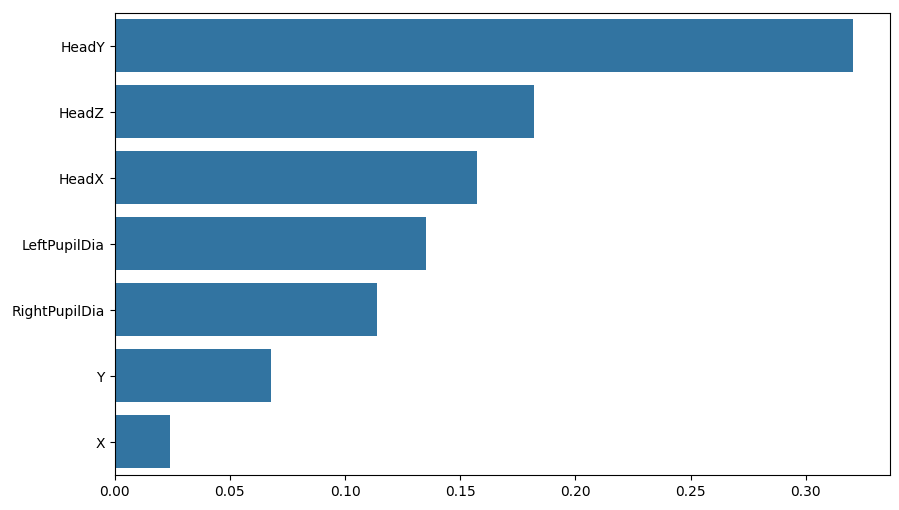}}
\caption{Visual analysis of student ID classification performance using raw gaze data under cognitive domain generalization (training on levels 1 and 2, testing on level 3): (a) Confusion matrix of the Random Forest model, highlighting strong identification performance for certain students (e.g., Student 6 and 7) and significant misclassification for underrepresented or overlapping gaze profiles (e.g., Student 1); (b) Confusion matrix of the Decision Tree model showing comparatively weaker generalization with greater class confusion, particularly for Student 7; and (c) Feature importance plot from Random Forest, showing that horizontal and vertical gaze coordinates contribute most significantly to student discrimination, indicating that spatial gaze behavior serves as a robust identity signature.}
\label{Fig.3}
\end{figure*}

The Random Forest confusion matrix in Fig.~\ref{Fig.3}(a) confirmed these trends, showing strong diagonal values for Students 6 and 7 and widespread misclassification for Students 1 and 5. The Decision Tree confusion matrix in Fig.~\ref{Fig.3}(b) showed slightly better balance for Student 9 (F1-score 0.58) but poorer performance for Student 7 (F1-score 0.59), highlighting its lower capacity to generalize complex gaze behaviors. The feature importance plot in Fig.~\ref{Fig.3}(c) revealed that horizontal and vertical gaze coordinates were the most influential predictors, underscoring the discriminative power of spatial gaze features in modelling student-specific visual attention. These results demonstrate that while raw gaze data can distinguish student identities with reasonable accuracy, model performance is sensitive to gaze consistency, task variation, and class imbalance, particularly in less-represented students.

\subsubsection{Scenario 3}
We evaluated the performance of classification models when the dataset was randomly split into 60\% training and 40\% testing. This represents an ideal, in-distribution evaluation setting without domain shift. The Random Forest classifier achieved a 5-fold cross-validation accuracy of 99.90\% and a test accuracy of 99.95\%, while the Decision Tree classifier reached 99.61\% on cross-validation and 99.73\% on the testing set. The classification reports are in the Table.~\ref{Table5} show nearly perfect performance across all classes: the Random Forest yielded precision, recall, and F1-scores of 1.00 for all students, and the Decision Tree showed only a slight drop for a few classes, such as Student 3 (recall = 0.99) and Student 5 (F1-score = 0.99). These results suggest that raw gaze features can provide highly accurate student identification when trained and tested on data drawn from the same distribution.

\begin{table}[!ht]
\centering
\caption{Random Forest and Decision Tree classification report for student ID prediction}
\label{Table5}
\begin{tabular}{llccc}
\toprule
\textbf{Model}&\textbf{Student ID} & \textbf{Precision} & \textbf{Recall} & \textbf{F1-score} \\
\midrule
\multirow{9}{*}{RF}&1 & 1.00 & 1.00 & 1.00 \\
&2 & 1.00 & 1.00 & 1.00 \\
&3 & 1.00 & 1.00 & 1.00 \\
&4 & 1.00 & 1.00 & 1.00 \\
&5 & 1.00 & 1.00 & 1.00 \\
&6 & 1.00 & 1.00 & 1.00 \\
&7 & 1.00 & 1.00 & 1.00 \\
&8 & 1.00 & 1.00 & 1.00 \\
&9 & 1.00 & 1.00 & 1.00 \\
\midrule
\textbf{Accuracy} & & & &\textbf{0.9995} \\
\midrule
\multirow{9}{*}{DT}&1 & 1.00 & 1.00 & 1.00 \\
&2 & 1.00 & 1.00 & 1.00 \\
&3 & 1.00 & 0.99 & 1.00 \\
&4 & 1.00 & 1.00 & 1.00 \\
&5 & 0.99 & 0.99 & 0.99 \\
&6 & 1.00 & 1.00 & 1.00 \\
&7 & 0.98 & 1.00 & 0.99 \\
&8 & 1.00 & 1.00 & 1.00 \\
&9 & 1.00 & 1.00 & 1.00 \\
\midrule
\textbf{Accuracy} & & && \textbf{0.9973} \\
\bottomrule
\end{tabular}
\end{table}

\begin{figure*}[!ht]
\centering
\subfloat[Random Forest\label{Fig.4(a)}]{
\includegraphics[width=0.49\textwidth]{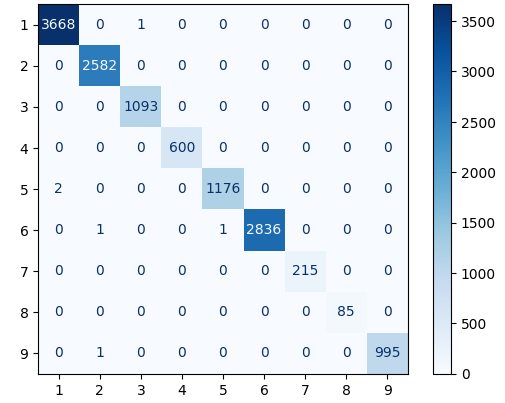}}
\subfloat[Decision Tree\label{Fig.4(b)}]{
\includegraphics[width=0.49\textwidth]{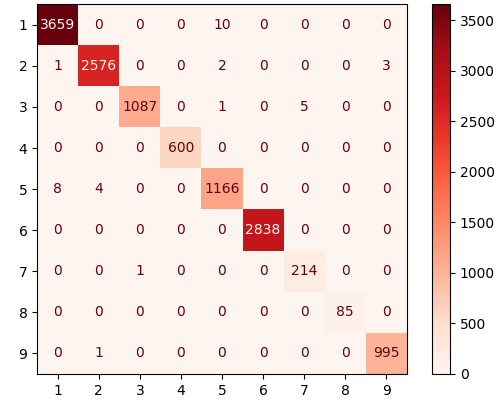}}\\
\subfloat[Feature Importance using Random Forest\label{Fig.4(c)}]{
\includegraphics[width=0.8\textwidth]{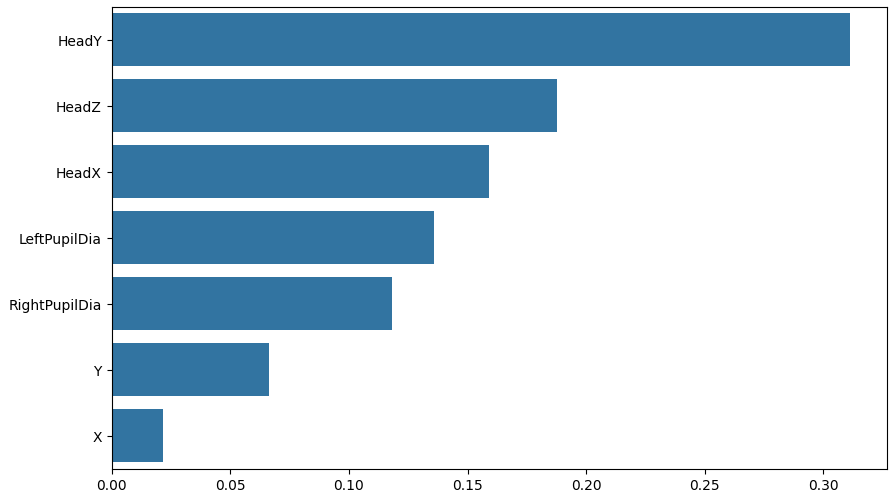}}
\caption{Model performance evaluation under idealized conditions with randomly partitioned data: (a) Confusion matrix of the Random Forest classifier showing perfect classification across all student IDs, with no off-diagonal errors; (b) Confusion matrix of the Decision Tree classifier, exhibiting near-perfect classification with only minor misclassifications (e.g., slight recall drop for Student 3 and Student 5); and (c) Feature importance plot from the Random Forest model indicating that spatial gaze coordinates (\textit{gaze\_x}, \textit{gaze\_y}) are the most dominant factors contributing to identity discrimination.}
\label{Fig.4}
\end{figure*}

The Random Forest confusion matrix in Fig.~\ref{Fig.4}(a) shows perfect predictions with all diagonal values matching their class support: e.g., 3669 correct predictions for Student 1, 2838 for Student 6, 1178 for Student 5, and 996 for Student 9 with zero misclassifications across all classes. The Decision Tree confusion matrix in Fig.~\ref{Fig.4}(b) is similarly accurate, with only marginal deviations: for example, Student 3 had 1087 correct predictions out of 1093, and Student 5 had 1164 correct predictions out of 1178, confirming the near-perfect recall and precision metrics. The feature importance plot in Fig.~\ref{Fig.4}(c) highlights gaze\_x and gaze\_y as the dominant features, followed by gaze derivatives or normalized gaze positions, indicating that spatial gaze behavior alone is sufficient for robust identification in low-variance settings. This scenario demonstrates the models' ability to memorize and reproduce student-specific gaze patterns with extremely high fidelity. However, such performance may not generalize under more realistic, cognitively varied conditions.

\subsubsection{Scenario 4}
\begin{figure}[!ht]
\centering
\subfloat[Elbow Method based Cluster Estimation\label{Fig5(a)}]{
\includegraphics[width=0.6\columnwidth]{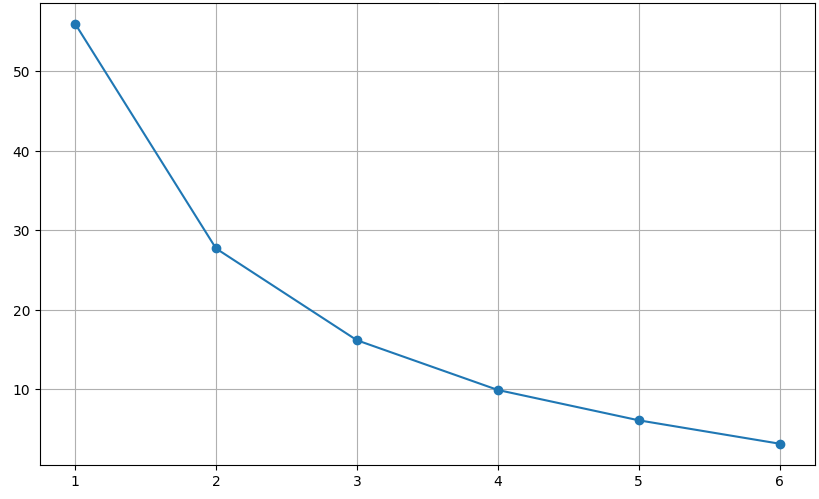}}\\
\subfloat[PCA \label{Fig5(b)}]{
\includegraphics[width=0.6\columnwidth]{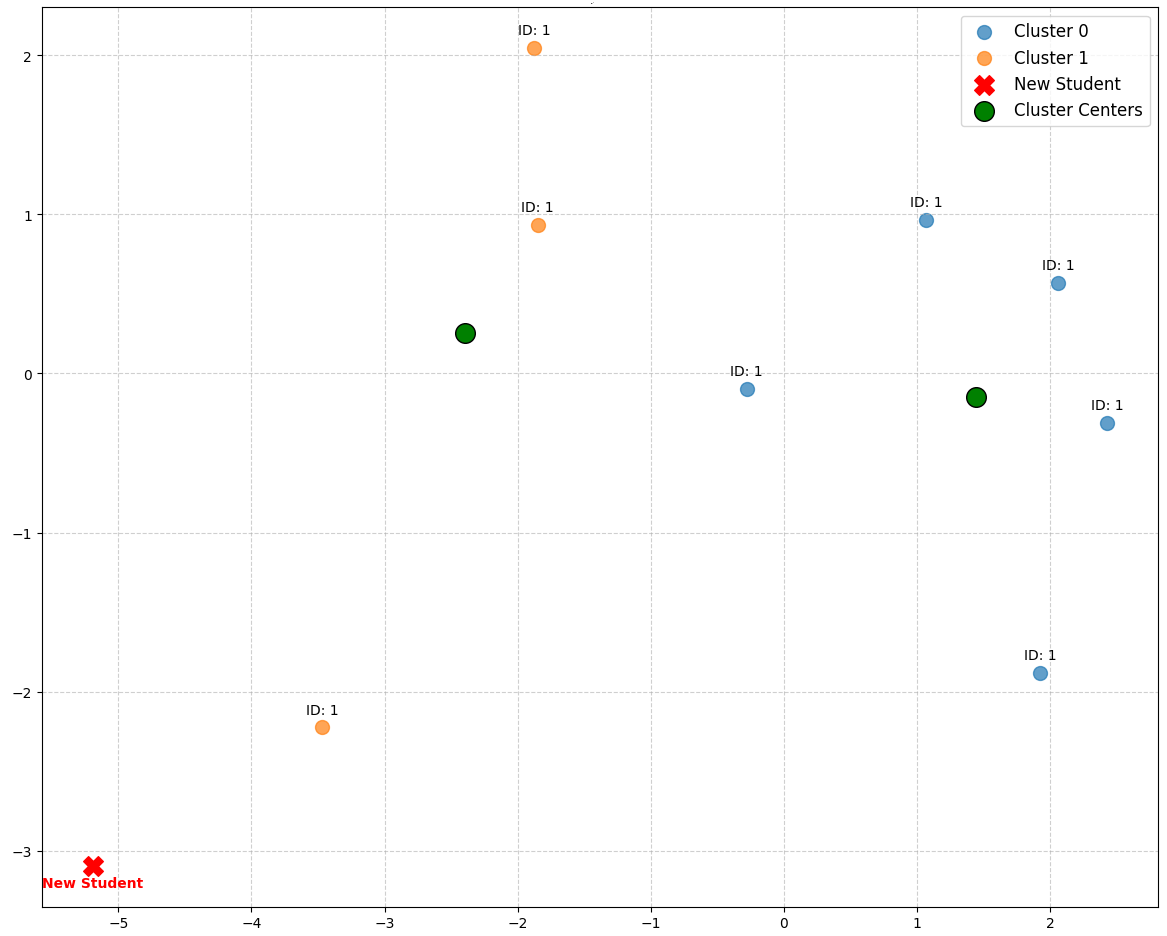}}
\caption{ (a) Determining the optimal number of clusters using the Elbow Method.(b) 2D visualisation of student feature space using Principal Component Analysis (PCA).}
\label{Fig5}
\end{figure}

In Scenario 4, we evaluated multiple unsupervised and rule-based approaches for detecting new, out-of-distribution students based on gaze features. The dataset consisted of eight known students, and a ninth student was introduced as a candidate for identity assignment. Across six approaches as shown in Fig. \ref{fig4}, the framework explored various methods: Approaches 1 and 3 immediately assigned a new ID (2) based on novelty detection logic. Approach 2 (Similarity Matching) and Approach 4 (KNN) identified the new student as the most similar to Student 1, but reported a similarity score of 2.69 and a confidence score of 0.0678, both of which are below the acceptance threshold, prompting the reassignment of a new ID. Approach 5 generated a feature-based encoded ID (7596) derived from the raw gaze values. This approach generates a hash from the feature string using MD5, and the very first 8 hex digits are converted to an integer ID (modulo 10,000 to keep it within range). If the ID already exists, it is incremented until a unique one is found, then printed along with the original feature string. Approach 6 employed K-Means clustering with outlier detection logic based on intra-cluster distances. The K-Means clustering plot in Fig.~\ref{Fig5}(a) illustrates the elbow method, where the x-axis represents the number of clusters (k), and the y-axis displays the within-cluster sum of squares (WCSS). The curve exhibited a distinct elbow around k = 2 to 4, suggesting a natural clustering structure in the known student data. Despite this, the ninth student's distance to the nearest cluster centre was 4.5446, exceeding the 95th percentile threshold of 2.5768, confirming that the student is an outlier and should be assigned a new identity. The PCA visualization in Fig.~\ref{Fig5}(b) provided an intuitive view of the clustering structure using the first two principal components, which captured 60.78\% (PC1) and 22.68\% (PC2) of the total variance. Each known student cluster was plotted in this 2D space, and the new student was marked with a red cross (\texttimes) isolated from all tight student clusters. This reinforced the conclusion that the new data point was not well-aligned with any existing identity, and unsupervised detection strategies correctly marked it as novel. The combined visual and quantitative results demonstrate how clustering, dimensionality reduction, and distance-based thresholds can be effectively integrated to manage open-set identification tasks in gaze-based frameworks.

\subsection{Phase 2 Results}
\begin{figure}[!ht]
 \centering
 \includegraphics[trim={0 0 0 1cm},clip,width=1\linewidth]{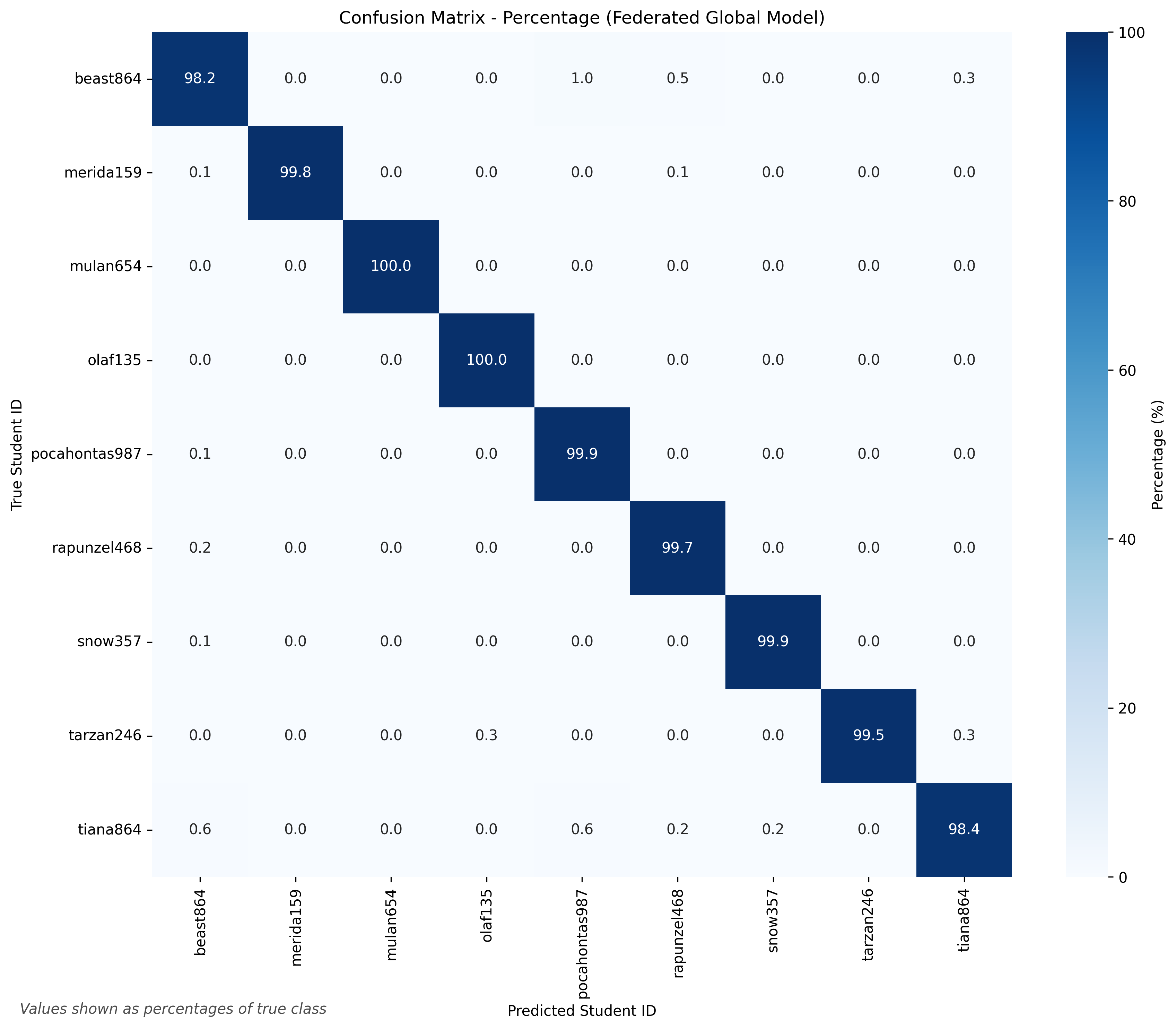}
 \caption{Confusion Matrix for Federated DNN Model}
 \label{figCM}
\end{figure}

Each square cell in Fig. \ref{figCM} contains a count of samples matching True ID and Predicted ID categories, which appear along the axes. The correctly classified instances appear within the diagonal matrix cells, whereas misclassification counts are located in the off-diagonal matrix cells. The matrix contains the anonymized identities of students. Furthermore, it shows that the proposed method correctly identified $mulan654$ and $olaf135$ in all instances. For the remaining classes, the model misclassified only a few instances each.

\begin{figure}[!ht]
 \centering
 \includegraphics[trim={0 0 0 0.6cm},clip,width=1\linewidth]{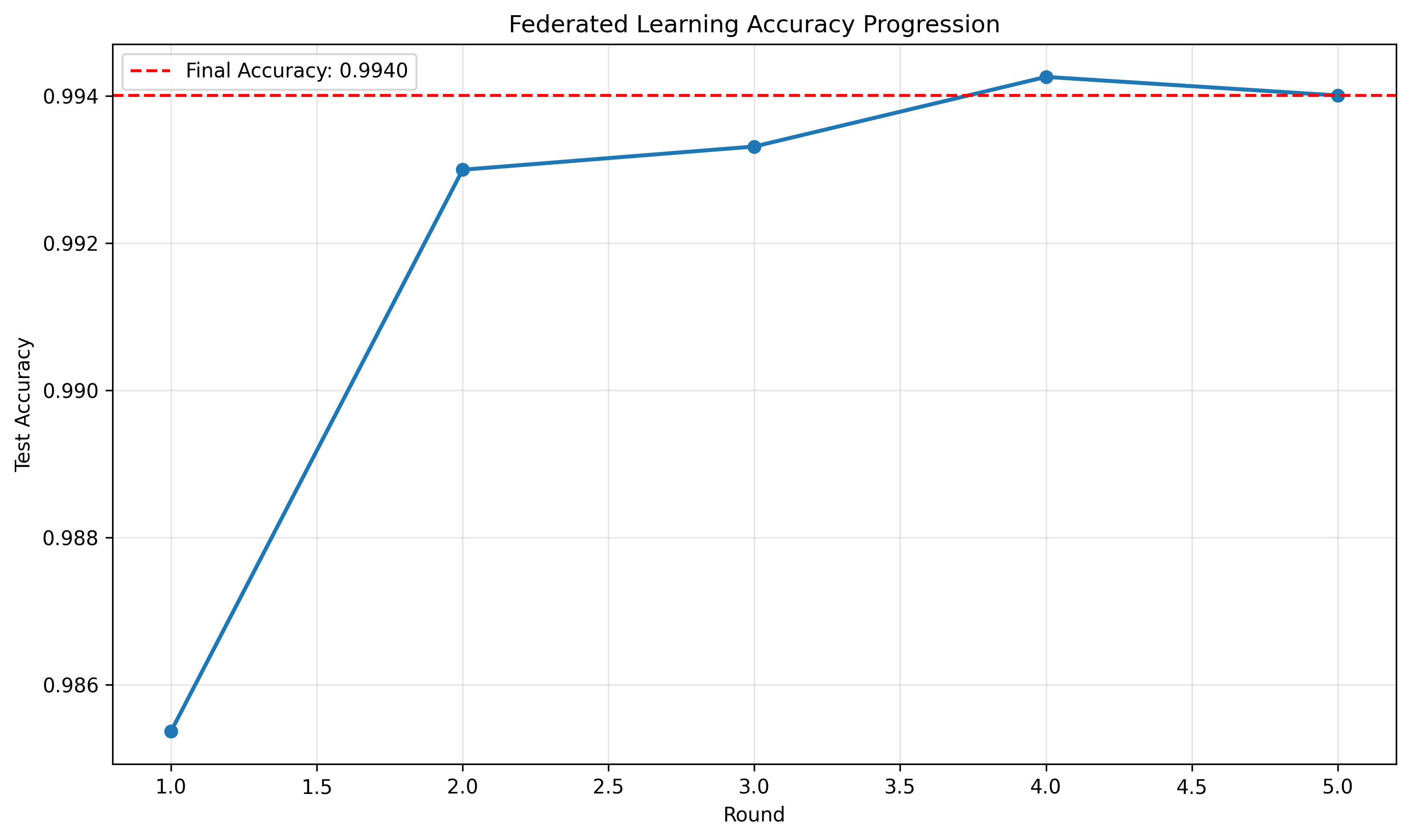}
 \caption{FL Progression Graph}
 \label{figACCPG}
\end{figure}

Fig. \ref{figACCPG} demonstrates the test accuracy development during five rounds in FL. It shows an initial accuracy growth from round 1 to round 5 (~0.98 to ~0.994). The accuracy almost stabilises in rounds 4 and 5. The red dashed line in the figure signifies the final accuracy level, which is 0.994. The reference mark compares accuracy variations during different rounds of the analysis. The model exhibits initial instability during updates, possibly due to unpredictable client contributions to the FL practice. Several rounds of iteration presented a stable model performance development after the initial fluctuation period.

\begin{figure*}[!ht]
 \centering
 \subfloat[Client 1]{
   \includegraphics[trim={0 0 0 1.35cm},clip,width=\linewidth]{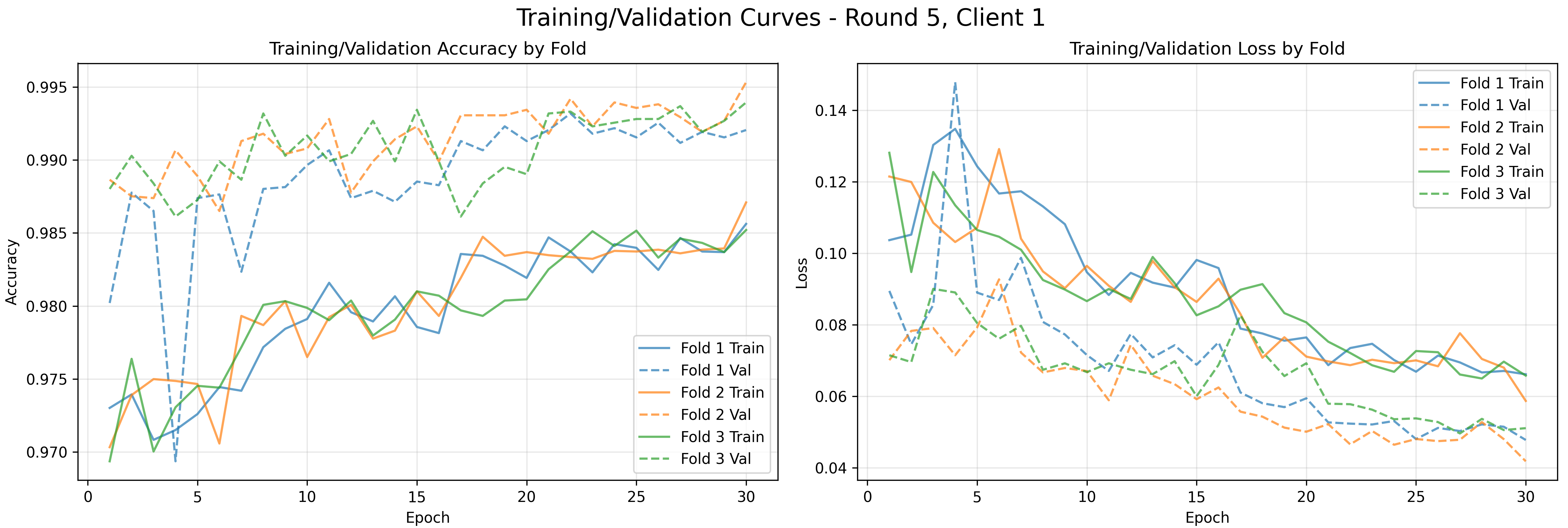} }
 \hfill
 \subfloat[Client 2]{
   \includegraphics[trim={0 0 0 1.35cm},clip,width=\linewidth]{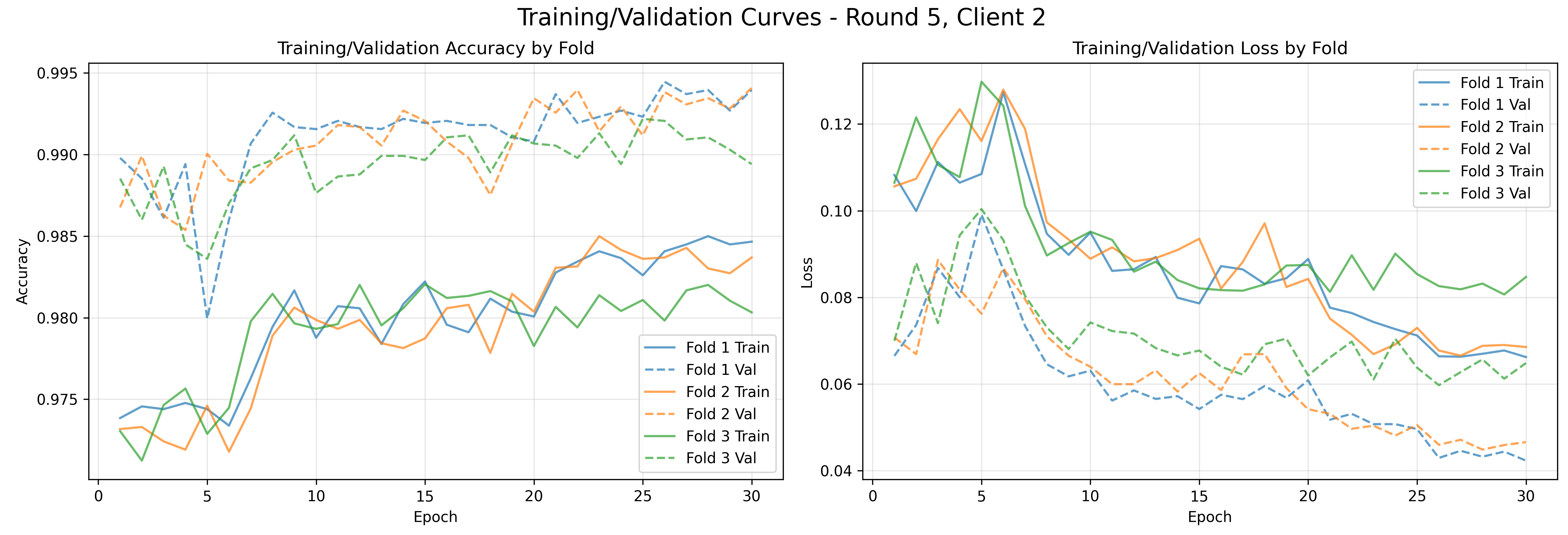}}
 \caption{Training and Validation curves for Clients in the final round (Round 5)}
 \label{TandV}
\end{figure*}

Figure \ref{TandV}(a) and Figure \ref{TandV}(b) show the training and validation curves for Clients 1 and 2 in the 5th Round across 3-fold cross-validation. The curves indicate a healthy learning pattern with minimal overfitting. In both figures, the accuracies for training and validation start around 97-98\% and gradually improve to 98-99\%, with validation accuracy often matching or even slightly exceeding training accuracy. The loss curves in both figures exhibit a consistent downward trend, starting around 0.06-0.12 and decreasing to 0.04-0.07, with both training and validation losses decreasing in tandem. This behavior suggests the model is generalizing well rather than overfitting. Overfitting would be indicated by training accuracy continuing to rise while validation accuracy plateaus or decreases, and training loss dropping significantly faster than validation loss. Instead, we observe that the validation metrics remain competitive with the training metrics throughout the epochs, indicating that the regularization techniques (dropout, L2 regularization, and early stopping) are effectively preventing overfitting, and the model is learning meaningful patterns that generalize to unseen data.

\section{Ethical Considerations}\label{ethical}
The pilot study complied with European ethical guidelines for research involving human subjects~\cite{academies2023european}. All legal guardians provided informed consent before participation, and the children's assent was obtained using language suitable for their age. Participants received comprehensive information regarding the study's objectives, the nature of the assignments, and their entitlement to withdraw at any time without repercussions. All gathered data were anonymised to preserve participant privacy and confidentiality, meaning no personally identifiable information remained. Care was taken to ensure comfort and minimize intrusiveness, as children were involved, and digital technologies, including eye-tracking systems and wearable devices, were utilized. The Mushroom Hunter game was thoughtfully created to be entertaining, developmentally appropriate, and devoid of any potentially upsetting material. Every data-gathering tool (such as the smartwatch and eye tracker) was non-invasive, and participants were given thorough instructions on how to use it before beginning. Before the study started, ethical approval was received from the research ethics committees of the University of Lisbon and Babeș-Bolyai University.

\section{Conclusion and Future Work} \label{conclusion}
This paper presented a novel two-stage privacy-preserving framework that prevents backtracking of students while enabling accurate diagnostic classification. The framework consists of two phases. In the first phase, it covered four scenarios: (I) It first demonstrated how to predict disorder diagnosis based on different game levels. II) How to predict student IDs based on different game levels? III) How can we predict student IDs based on randomised data? IV) How can K-Means be used for out-of-sample data (treating the first 8 students as existing and the 9th as new)? In the second phase, we presented a two-stage privacy-preserving framework that prevents participants from being tracked back while enabling diagnostic classification. Further, we utilize FL across multiple clients, incorporating a secure identity management system with dummy IDs and administrator-only access controls. The proposed framework achieved 99.3\% accuracy using a decision tree for scenario 1 in phase 1; for scenario 2, it achieved an accuracy of 63\% using RF; for scenario 3, it achieved an accuracy of 99.7\% using a decision tree; for scenario 4, it successfully identified and assigned a new student ID, for the last scenario, it achieved an overall accuracy of 99.40\%. For Phase 2, it successfully prevented backtracking and secured an identity management system with dummy IDs and administrator-only access controls, achieving an overall accuracy of 99.40\%. In the future, we intend to collect a large dataset from NDD children with various disorders and improve the accuracy of backtracking. Further, we intend to provide a fast and robust framework.

% \subsection{CRediT authorship contribution statement}
% Abdul Rehman: Conceptualization, Writing – original draft, Writing – review \& editing, Visualization, Methodology, Analysis and/or interpretation of data, Formal analysis, Data curation. 
% Are Dæhlen: Writing – original draft, Review \& editing, Visualization, Methodology, Formal analysis.
% Ilona Heldal: Writing – original draft, Review \& editing, Supervision,  
% Jerry Chun-Wei Lin: Writing – original draft, Writing – review \& editing, Supervision.

\section*{Data availability}
Data will be made available on request.

\section*{Declaration of competing interest}
The authors share no conflict of interest.

\subsection{Acknowledgment}\label{ack}
The research leading to these results is within the frame of the ”EMPOWER. Design and evaluation of technological support tools to empower stakeholders in digital education” project, which has received funding from the European Union’s Horizon Europe programme under grant agreement No 101060918. Views and opinions expressed are, however, those of the author(s) only and do not necessarily reflect those of the European Union. Neither the European Union nor the granting authority can be held responsible for them.

\bibliographystyle{ieeetr}
\bibliography{ref-privacy}

\begin{IEEEbiography}{Abdul Rehman} is a Ph.D. candidate in Computer Science at the Department of Computer Science, Electrical Engineering and Mathematical Sciences, Western Norway University of Applied Sciences, Bergen, Norway. His research interests include privacy-preserving machine learning, eye tracking, human-computer interaction, and educational technologies.
\end{IEEEbiography}

\begin{IEEEbiography}{Are Dæhlen} is a Ph.D. candidate at the Department of Computer Science, Electrical Engineering and Mathematical Sciences, Western Norway University of Applied Sciences, Bergen, Norway. His research focuses on artificial intelligence in education, learning analytics, and user privacy in data-driven systems.
\end{IEEEbiography}

\begin{IEEEbiography}{Ilona Heldal} is a Professor at the Department of Computer Science, Electrical Engineering and Mathematical Sciences, Western Norway University of Applied Sciences, Bergen, Norway. She holds a Ph.D. in technology and has extensive experience in virtual reality, human-computer interaction, and collaborative learning environments.
\end{IEEEbiography}

\begin{IEEEbiography}{Jerry Chun-Wei Lin} is a Professor at the Department of Computer Science, Electrical Engineering and Mathematical Sciences, Western Norway University of Applied Sciences, Bergen, Norway. He has authored over 500 research articles and is recognized for his contributions to data mining, deep learning, and privacy-preserving algorithms. He serves on the editorial boards of multiple international journals.
\end{IEEEbiography}

\EOD
\end{document}